\documentclass{amsart}
\usepackage{adjustbox}
\usepackage[english]{babel}
\usepackage[T1]{fontenc}
\usepackage{url}
\usepackage{xcolor}

\newcommand{\bq}{\textbf{q}}

\renewcommand{\le}{\leq}

\renewcommand{\to}{\rightarrow}

\newcommand{\fer}[1]{(\ref{#1})}
\def\be#1\ee{\begin{equation}#1\end{equation}}
\newenvironment{equations}{\equation\aligned}{\endaligned\endequation}
\newcommand{\R}{\mathbb R}

\def\fo{\widehat f}

\def \WWD{\mathcal{W}}
\def \WFD{\mathcal{F}}
\def \WGD{\mathcal{G}}
\def\be#1\ee{\begin{equation}#1\end{equation}}

\renewcommand{\theequation}{\arabic{section}.\arabic{equation}}
\renewcommand{\thetable}{\arabic{section}.\arabic{table}}
\renewcommand{\thefigure}{\arabic{section}.\arabic{figure}}
\def\bqa{\begin{eqnarray}}
\def\eqa{\end{eqnarray}}

\def\bC{{\bf C}}

\newcommand{\bd}{\begin{displaymath}}
\newcommand{\ed}{\end{displaymath}}
\newcommand{\ba}{\begin{eqnarray}}
\newcommand{\ea}{\end{eqnarray}}

\def\fo{\widehat f}

\def\R{\mathbb{R}}

\def\bx{{\bf x}}
\def\bX{{\bf X}}
\def\bY{{\bf Y}}

\def\by{{\bf y}}
\def\bm{{\bf m}}

\def\bxi{{\pmb\xi}}
\def\bieta{{\pmb\eta}}

\def \PP{\mathcal{P}}

\def \erre{\mathbb{R}}

\usepackage{hyperref}
\usepackage{booktabs}
\usepackage[inline]{enumitem}
\usepackage{verbatim}

\newtheorem{theorem}{Theorem}
\newtheorem{proposition}[theorem]{Proposition}

\newtheorem{corollary}[theorem]{Corollary}
\newtheorem{remark}[theorem]{Remark}

\newtheorem{definition}[theorem]{Definition}

\begin{document}

\markboth{G. Auricchio, G. Brigati, P. Giudici, G. Toscani}{Multivariate Gini-type discrepancies}

%
%

\title{MULTIVARIATE GINI-TYPE DISCREPANCIES}


\author[G.~Auricchio]{Gennaro Auricchio}
\address{Gennaro Auricchio: Department of Mathematics, University of Padua, via Trieste 36,
Padova, 35100 Italy}
\email{gennaro.auricchio@unipd.it}

\author[G.~Brigati]{Giovanni Brigati}
\address{Giovanni Brigati: Institute of Science and Technology Austria, Am Campus 1, Klosterneuburg, 3400 Austria}
\email{giovanni.brigati@ist.ac.at}

\author[P.~Giudici]{Paolo Giudici}
\address{Paolo Giudici: Department of Economics and Management, University of Pavia, 27100 Italy}
\email{paolo.giudici@unipv.it}

\author[G.~Toscani]{Giuseppe Toscani}
\address{Giuseppe Toscani: Department of Mathematics, University of Pavia, and IMATI CNR, via Ferrata 1, Pavia, 27100 Italy}
\email{giuseppe.toscani@unipv.it}



\begin{abstract}
Measuring distances
in a multidimensional 
setting is a challenging problem, which appears in many fields of science and engineering.  In this paper, to measure the distance between two multivariate distributions,  we introduce a new measure of discrepancy which is scale invariant and which, in the case of two independent copies of the same distribution, and after normalization,  coincides with the scaling invariant multidimensional version of the Gini index recently proposed in~\cite{giudici2024measuring}.  A byproduct of the analysis is  an easy-to-handle  discrepancy  metric, obtained by application of the theory to a pair of Gaussian multidimensional  densities. The obtained metric does improve the standard metrics, based on the mean squared error, as it is scale invariant. The importance of this theoretical finding is illustrated by means of a real  problem that concerns  measuring the importance of Environmental, Social and Governance factors for the growth of small and medium enterprises.  
\end{abstract}
\vskip.2cm


\maketitle




\section{Introduction}
In this paper we are interested in introducing some \emph{distance} or \emph{discrepancy} between  statistical distributions, which consist of  $n$-component vectors, $n >1$, whose components  can be measured  with respect to different units of measure,  obtained from each other  by multiplication with a positive constant. This problem is closely related to measure the heterogeneity of a single distribution, which consists of a $n$-component vector,  and allows to better understand the nature of the 
variability
expressed by a multidimensional distribution. 

The interest in measuring the 
heterogeneity of  statistical distributions arises in many fields of science and engineering, and it is one of the fundamental features of statistical analysis.\cite{HR} Originally, one-dimensional heterogeneity measures have been designed in connection with problems in the field of economics, 
for which the interest was not to measure the variability of a set of observations from a mean but, rather, to measure their mutual variability, with the final aim of quantifying the inequality in the distribution of income, wealth, or consumption.

In this context, the most used measure of inequality is the Gini index, first proposed by the Italian statistician Corrado Gini more than a century ago,\cite{Gini1,Gini2} along with the less well known index proposed by Gaetano Pietra.\cite{Pie} The interest in inequality measures in economics is still alive, as documented by those introduced in Refs. \cite{GudRaf2024,BL,Cou,Cow,GudRaf2023,HN}. For an exhaustive review of the state of the art about inequality measures  we refer to the recent contributions \cite{Ban,Eli,Eli3}. 

As a matter of fact, there is no universally accepted definition of inequality and/or heterogeneity of a statistical distribution, even if the various \emph{one-dimensional} proposals are characterized by some \emph{universal properties}. Most definitions agree with the property that a  distribution with the whole mass concentrated in a single point, and all others points with mass equal to zero, is the most heterogeneous and unequal distribution. Likewise, most agree with the fact that a distribution with all its mass  spread evenly over all points is the least heterogeneous and unequal distribution. As observed in Ref. \cite{HR}, it seems further reasonable to assume that any \emph{good} measure of heterogeneity should increase as the mass moves toward one with all mass in one point. Similarly, heterogeneity should decrease when the mass becomes more evenly distributed. \\ Moreover, if we double the mass in the two mentioned extreme distributions, the former still has only one point with all the mass and the latter  still has its mass evenly distributed. Thus,  it is highly reasonable to consider that a measure of heterogeneity should be invariant under multiplication by a constant, as under doubling. 
We will refer to this last property of a heterogeneity measure as the \emph{scaling invariance} property.
Likewise, it is natural to assume that adding a constant to 
a distribution decreases its heterogeneity.  As observed in Ref. \cite{HR},
this is highly intuitive.  In the income setting,  giving to each individual a certain fixed 
amount of income  will have the effect to decrease the inequality of the distribution. Having in mind  income distributions, we will refer to this property as the \emph{uniform redistribution} property.
\\
Despite the enormous amount of research concerned with the application of heterogeneity measures, and of the Gini index in particular, very few studies consider the case of a  multidimensional distribution, which consists of a $n$-component vector, $n >1$. 
Among them, the most recent contribution Ref.
\cite{Grothe} provided the expression of a bivariate Gini coefficient by means of a  copula-based approach, whereas Ref. \cite{Sarabia} exploited the notion of a Lorenz surface, an extension of the univariate Lorenz curve\cite{Lor} to higher dimensions, to obtain  a generalized Gini index, expressed as a function of marginal Gini 
indices. 

The aim of obtaining a multivariate Lorenz curve, which could lead to a multivariate Gini index, was previously attempted in Refs. \cite{Arnold,KoshMos96,Tagushi1,Tagushi2} and Ref. \cite{KoshMos97}, who introduced the concept of a Lorenz Zonoid.
While theoretically sound, as representing the mathematical and geometrical extension of the one-dimensional Gini coefficient, Lorenz Zonoids are difficult to implement in practice, especially for computational reasons. Moreover, they do not seem to have properties similar to those previously mentioned for unidimensional measures.\cite{Decancq} For instance, the \emph{scaling invariance} property, which holds true in the one-dimensional case, is typically not covered by the multidimensional Lorenz Zonoids. This is a relevant weakness, as a measure which is not scale invariant may lead to different rankings of the considered observations, when the scale of one or more variables is changed.\\
Likewise, in the multidimensional setting, it seems difficult to verify the \emph{uniform redistribution} property.

To address these shortcomings, new multidimensional inequality measures with the scaling invariance property were considered in Ref. \cite{giudici2024measuring}, first by resorting to the Fourier transform of a probability distribution, and subsequently by extending the same strategy to obtain new multivariate expressions of both Gini and Pietra indices. These new definitions possess the \emph{scaling invariance} and the \emph{uniform redistribution} properties, and can be obtained 
through substitution of the Euclidean metric, 
on which the one-dimensional indices are based, with the Mahalanobis distance.\cite{Maha}

Following the strategy in Ref. \cite{giudici2024measuring}, in this paper we will propose scaling invariant discrepancies between multidimensional distributions which possess both the \emph{scaling invariance} and the \emph{uniform redistribution} properties along the components of each distribution. 
After the mathematical specification of the properties required to a multidimensional inequality distribution in Section \ref{sec:properties}, in Section \ref{sec:transport}, we shall see that this problem is closely connected with solving a new scaling invariant mass transport problem.  In  Section \ref{sec:Fourier} we will propose  a new scaling invariant discrepancy measure expressed in terms of the Fourier transform. The consistency among the discrepancies in Sections \ref{sec:transport} and \ref{sec:Fourier} will be assessed looking at their explicit expression, when evaluated in correspondence to multidimensional Gaussian distributions. The resulting easy-to-handle expression can be fruitfully used to have a quick evaluation of the discrepancy between two distributions in terms of their means and covariance matrices. 
We conclude the paper with a practical application to the important problem of assessing the safety of artificial intelligence applications in Section \ref{sec:examples}.
In Table \ref{table:notation}, we report the notation we use throughout the paper.

\begin{table}[!ht]
    \centering
    \begin{adjustbox}{width=0.9\textwidth}
    
    \footnotesize
    \begin{tabular}{ll}
        \toprule
        
        Notation & Description \\ 

        \hline
        
        $\bX\sim\mu$ & Random vector with associated law $\mu$ \\ 
        $\mathbb{E}[\bX]$ & Expected value of the random vector $\bX$ \\ 
        $\Sigma_\mu$ & Covariance matrix associated with $\mu$ \\  
        $P_\mu$ & Correlation matrix associated with $\mu$ \\
        $|\bx|$ & Euclidean norm of the vector $\bx$ \\
        $\PP(\erre^n)$ & Set of probability measures over $\erre^n$ \\
        $\PP_2(\erre^n)$ & Set of probability measures over $\erre^n$ with finite second moment \\
        $W_\mu^{ZCA}$ & ZCA-correlation matrix associated with the probability law $\mu$ \\
        $\bX^*\sim\mu^*$ & ZCA-correlation whitened random vector obtained from $\bX\sim\mu$ \\
        $\Pi(\mu,\nu)$ & Set of transportation plan between $\mu$ and $\nu$ \\
        $\WWD$ & Whitened Wasserstein discrepancy function \\
        $\WFD$ & Whitened Fourier discrepancy function \\
        $\WGD$ & Gini Discrepancy function \\
        
        \bottomrule
    \end{tabular}
    \end{adjustbox}
    \caption{Summary of the main notation used throughout the paper.}
    \label{table:notation}
\end{table}

\section{Preliminaries}

In what follows, we denote vectors with bold letters and use standard letters for scalar quantities.
Let $P_2(\R^n)$ denote the class of all probability measures $\mu$ on  $\R^n$  such that
\begin{equation}
    \label{eq:secondmoment}
    \int_{\R^n} |\bx|^2 \mu(d\bx) < + \infty,
\end{equation}
where we denote with $|\bx|$ the Euclidean norm of $\bx$, that is
\begin{equation}
|\bx|=\sqrt{x_1^2+x_2^2+\dots+x_n^2}.
\end{equation}
Condition \eqref{eq:secondmoment} ensures that the mean value vector  $\bm_\mu := \mathbb{E}(\bX)= \int_{\R^n} \bx \, \mu(d\bx)$ and the covariance matrix $\Sigma_\mu$, namely
\[
(\Sigma_\mu)_{ij} = cov(X_i,X_j) := \int_{\R^n} x_i \, x_j \, \mu(d\bx) - \left( \int_{\R^n} x_i \mu(d\bx) \right) \, \left( \int_{\R^n} x_j \mu(d\bx) \right),
\]
of a random vector ${\bf X}$, whose probability law is $\mu \in P_2(\R^n)$, are well defined. 
Similarly, the correlation matrix of $\mu$, namely $P_\mu$, is well-defined and we have that
\[
(P_\mu)_{i,j}=\frac{(\Sigma_\mu)_{i,j}}{\sqrt{Var(X_i)Var(X_j)}}.
\]
Since we only work on the states space $\mathbb R^n$, we identify each random vector $\bX$ with its law $\mu$. Henceforth, we use $\bX$ and $\mu$ interchangeably and write $\bX\sim\mu$.
We remark that $\Sigma_\mu$ and $P_\mu$ are symmetric and positive semidefinite.
In what follows, we assume that $\Sigma_\mu$ is invertible, unless we specify otherwise.

\subsection{The Whitening Process}\label{sec:SI}

Given a $n$-dimensional random  vector $\bX$, of mean $\mathbb{E}(\bX)= {\bf m} = (m_1,m_2,\dots, m_n)^T$ and a positive definite  $n\times n$ covariance matrix $\Sigma$, a whitening  process on $\bX$ returns a new  $n$-dimensional random vector 
\begin{equation}
    \label{whi1}
{\bX}^{*} = W{\bf X}
\end{equation}
whose covariance is the identity matrix $I$.  
Any $n\times n$ square matrix $W$ which yields this result is called the whitening matrix.
The entries of $\bf X^*$ are known as the \textit{principal components} of $\bf X$.
The condition $cov(X^*_i,X^*_j)= 0$ for $i\not=j$, the multivariate  analysis of the random vector $\bX^*$ simplifies both from a computational and a statistical viewpoint.
For this reason, whitening is a critically important tool, which is often employed for data pre-processing in statistics and machine learning applications.\cite{ex1,ex2}
The whitening transformation defined in \fer{whi1} requires that the whitening matrix $W$ satisfies $W\Sigma W^T = I$, which implies $W (\Sigma W^TW) = W$.
Therefore, $W$ has to satisfy the identity

\begin{equation}
\label{whi2}
W^TW =\Sigma^{-1}.
\end{equation}

It is important to outline that, given a $n$-dimensional random vector $\bX$ whose covariance matrix is $\Sigma$, condition \fer{whi2} does not fully identify $W$, but allows for rotational freedom. 
Indeed, if we set
\[
W = Z\Sigma^{-1/2},
\]
where $Z$ is an orthogonal matrix so that $Z^TZ = I$, $W$ satisfies \fer{whi2} regardless of the choice of $Z$. 
Owing to the fact that the whitening matrix is not unique, there are a variety of whitening processes that are commonly used.\cite{LZ} 
However, as shown in Ref. \cite{AGT},  only a few of them possess the \emph{scale stability} property, which ensures that the random vector obtained by whitening $\bX:=(X_1,X_2,\dots,X_n)$ or $\bX'=(X_1,X_2,\dots,aX_i,\dots,X_n)$ is the same for every $a>0$ and $i\in[n]$.
%
More formally, a whitening process is a map $\mathcal{S}$ that, given in input a probability distribution $\mu$ or a random vector $\bX\sim\mu$, returns a whitening matrix $W_\mu$.
Denoted with $diag(\vec q)$ the diagonal matrix whose diagonal values are $\vec q =(q_1,q_2,\dots,q_n)$, we have that $\mathcal{S}$ is scale stable if, given a random vector $\bX$ and a diagonal matrix $Q=diag(q_1,q_2,\dots,q_n)$ such that $q_i \geq 0$, it holds
\[
    W_\mu\bX=W_{\mu^{Q}}(Q\bX),
\]
where $\mu^Q$ is the probability measure associated with the random vector $Q\bX$.

A \emph{Scale stable} whitening processes exist; for example, in Ref. \cite{AGT}, it has been shown that the Cholesky whitening and the ZCA-cor whitening transformation are both Scale Stable.
For the sake of simplicity, we focus only on the ZCA-cor whitening associated with a probability measure $\mu$, which is defined as
\begin{equation}
    \label{ZCA}
W_\mu^{ZCA} = P_\mu^{-1/2}V^{-1/2}=G^T\Theta GV^{-1/2},
\end{equation}
where 
\begin{enumerate}
    \item $V$ is a diagonal matrix containing the variances of the components of $\bX$,
    \item $G$ is the orthonormal matrix induced by the eigenvectors of $P_\mu$, and
    \item $\Theta$ is the diagonal matrix containing the eigenvalues of $P_\mu$.
\end{enumerate} 
It is essential to remark that not all whitening processes possess the property of \emph{scale stability}. 
For example the PCA whitening does not enjoy such property.\cite{AGT}

\subsection{Transport and Optimal Transport}\label{ss:OT}

Over the last two decades, Optimal Transport has became a standard way to define metrics to compare how similar two probability measures are. \cite{villani2009optimal,santambrogio2015optimal,ambrosio2005gradient,peyre2019computational,cuturi2013sinkhorn} 
Given two probability measures, namely $\mu$ and $\nu$, we consider the set of couplings between $\bX\sim \mu$ and $\bY\sim \nu$.
Each coupling, is associated with a joint probability distribution which is called \textit{transportation plan}.
The set of transportation plans between two probability measure $\mu$ and $\nu$ is thus defined as
\[
\Pi(\mu,\nu) := \{ \pi \in \mathcal{P}(\R^{2n}) \, : \, (e_x)_\# \pi = \mu, \quad (e_y)_\# \pi = \nu \},
\]
where $e_x:\R^{2n}\to\R^n$ and $e_y:\R^{2n}\to\R^n$ are defined as it follows
\[
e_x (\bx,\by) = \bx, \qquad e_y(\bx,\by) = \by,
\]
and $F_\#\mu$ is the pushforward of a measure through a measurable map $F:\R^n\to\R$, which is given by the following identity
\[
\int_{\R^n} \phi(\by) \, F_\# \mu(d\by) = \int_{\R^n} \phi(F(\bx)) \, \mu(d\bx), 
\] 
which has to be satisfied for any test function $\phi:\R^n\to\R$. 
Notice that any couple of measures $\mu,\nu$ induces several transportation plans $\pi$, depending on how much correlated the two marginals are.
For the sake of our discussion, we consider two types of transportation plans:
\begin{enumerate}
    \item[\textbf{1.}] If $\mu$ and $\nu$ are assumed to be independent, the probability law associated to the coupling is 
    \[
    \pi(d\bx,d\by) = \mu(d\bx) \, \nu(d\by) =: \mu \otimes \nu. 
    \]
    According to $\pi=\mu \otimes \nu$, the proportion of pairs of the type $(\bx,\by)$ is the product of the proportion of data items of the kind $\bx$ in $\mu$ times the proportion of data items of the kind $\by$ in $\nu$.
    In other words, the data items of the type $\bx$ from $\mu$ are uniformly associated to all types of data from $\nu$, for any type $\bx$.
    \item[\textbf{2.}] Assume now that there exists a measurable function $T:\R^n\to\R^n$ such that $T_\#\mu=\nu$. 
    In this case, the probability measure $\pi_T:=(Id,T)_\#\mu$ is a transportation plan between $\mu$ and $\nu$.
    In particular, $\pi_T$ is the probability measure associated with a coupling $(\bX,\bY)$ such that $\bY=T(\bX)$, i.e. one of the two variables can be expressed as a function of the other.
\end{enumerate}

\begin{remark}
    The transportation plans described in \textbf{1.} and \textbf{2.} are two extremes.
    The plan $\mu \otimes \nu$ represent the fact that $\mu$ and $\nu$ are assumed to be independent, hence there is no correlation between $\bX$ and $\bY$.
    The most different transportation plan is the one describing a full \emph{functional dependence} between $\bX$ and $\bY$, i.e. $\pi_T:=(Id,T)_\#\mu$, as this case yields the highest correlation.
\end{remark}

Given a cost function $c : \R^n\times\R^n \to \mathbb R_+$, the optimal transport problem consists in selecting the transportation plan between $\mu$ and $\nu$, that minimizes the following cost functional over $\Pi(\mu,\nu)$
\[
\mathsf c(\pi) = \int_{\R^n\times\R^n} c(\bx,\by) \, \pi(d\bx,d\by).
\]
When $c(\bx,\by) = |\bx-\by|^p$, the minimum of $\mathsf c$ induces the $p$-th Wasserstein distance $W_p$, so that
\begin{equation}
    \label{wdp}
W^p_p(\mu,\nu)=\inf_{\pi\in\Pi(\mu,\nu)}\int_{\R^n\times\R^n} |\bx-\by|^p \, \pi(d\bx,d\by).
\end{equation}
For $p=2,$ we have the classical Wasserstein distance, where \emph{far away points} are disadvantaged. 
In this paper, we focus on the $1$-Wasserstein distance.
We refer the reader to Ref. \cite{ambrosio2005gradient} for a comprehensive discussion of Wasserstein distance and the many deep mathematical properties they possess.
A striking result by Yann Brenier,\cite{brenier1991polar} showed that, under mild assumptions on $\mu$, the infimum of \eqref{wdp} is attained on a coupling induced by a transport map $T$.
So that the optimal transportation plan is given by $\pi_{opt}=(Id,T)_\# \mu$, where $Id$ is the identity map and $T$ is the \emph{optimal transport map} between $\mu$ and $\nu$. 
Heuristically, the fact that optimality is attained on a plan induced by a transport map can be explained as follows. 
Assume that coupling an item of the type $\bx$ to an item of the type $\by$ has a particularly convenient cost $c(\bx,\by)$. 
Then, an optimal transport plan would privilege pairs $(\bx,\by)$ as much as possible, instead of associating the type $\bx$ to any other class of data from $\nu$. 
When it is possible to couple all items $\bx$ only with items of the type $\by$, we write $\by=T(\bx)$. 
%
%
As said above, the plan $\mu \otimes \nu$ goes exactly in the converse direction, as any item $\bx$ from the set $\mu$ is assigned to all items $\by$ from $\nu$. 
In brief, \emph{optimal transport prefers maximal correlation}, see Ref. \cite{brenier1991polar}, and later Ref. \cite{evans1999differential}. 
To make an example, assume that $\mu \sim \mathcal{N}(\bm_1,\Sigma_1)$ and $\nu \sim \mathcal{N}(\bm_2,\Sigma_2)$, \emph{i.e.}~both measures are distributed according to a Gaussian distribution.
In this case, there is an explicit, and affine, map $T$ inducing the optimal transport coupling of \eqref{wdp} when $p=2$, given by
\begin{equation}
    \label{otg}
    \by = T(\bx) := \bm_2 + A (\bx-\bm_1), \qquad A = \Sigma_1^{-1/2} \, ( \Sigma_1^{1/2} \, \Sigma_2 \, \Sigma_1^{1/2}  )^{1/2} \Sigma_1^{-1/2},
\end{equation}
therefore
\begin{equation}
    \label{otg2}
    W_2^2(\mu,\nu) = |\bm_1-\bm_2|^2 + \mathrm{Tr}( \Sigma_1+\Sigma_2 -2 (\Sigma_1^{1/2} \, \Sigma_2 \,\Sigma_1^{1/2})^{1/2}), 
\end{equation}
see\cite{chafai2010} and the references quoted therein. Note that the second term vanishes whenever $\Sigma_1 = \Sigma_2$, and equals to $\mathrm{Tr}((\Sigma_1-\Sigma_2)^2)$ if the matrices $\Sigma_1$ and $\Sigma_2$ commute.

\begin{remark}
Despite their appealing mathematical properties, the Wasserstein Distances are often hindered by their high computational cost.
For this reason, many easy-to-compute metrics that are equivalent to the Wasserstein Distance have been studied.
Among the many, we focus on the Fourier Based Metrics, which were firstly introduced to study the trend to equilibrium for solutions of the spatially homogeneous Boltzmann equation for Maxwell molecules.\cite{GTW}
Given $\mu \in \mathcal P_2(\R^n)$, let $\fo(\bxi)$ for $\bxi  \in \R^n$ denote the Fourier transform of $\mu$, so that
\[
\fo(\bxi) =  \int_{\R^n } e^{-i \bxi^T\bx} \, \mu(d\bx).
\]
The Fourier Based Metric between $\mu$ and $\nu$ is then defined as
\begin{equation}
    \label{fm1}
d_1(\mu,\nu) = \sup_{\bxi \in \R^n} \frac{|\hat f(\bxi) -\hat g(\bxi))|}{|\bxi|},
\end{equation}
where $\hat f$ (respectively $\hat g$) are the Fourier transforms of $\mu$  (respectively $\nu$). 
It was shown in \cite{CaTo} and later in \cite{Au1,Au2}, that the Fourier-based metric $d_1$ are equivalent to Wasserstein Distances.
In what follows, we will focus mostly on Wasserstein Distances, however, due to their equivalence, the same arguments and techniques can be adopted to derive similar results for \eqref{fm1}.
Moreover, in Ref. \cite{To1} it has been shown that these metrics can be used to define sparsity indexes for multivariate random vectors. 
\end{remark}

\subsection{From Inequality properties to discrepancy measures}\label{sec:properties}

Following the line of thought of Ref. \cite{HR}, we search for measures of discrepancy between two multidimensional distributions that possess both the \emph{scaling invariance} and the \emph{uniform redistribution} properties.

\subsubsection{The Scaling Invariance Property}

Let \[{\bf X} =(X_1,X_2, \dots X_n)^T, \qquad{\bf Y}=(Y_1,Y_2, \dots Y_n)^T\]
denote two random vectors in $\R^n$.
Moreover, let us denote with $\mu$ and $\nu$, the probability measures associated with $\bX$ and $\bY$, respectively.

\begin{definition}[Scaling invariant property]
Let $\delta:\PP(\R^n)\times \PP(\R^n)\to [0,\infty)$ be a discrepancy.
Then $\delta$ is \emph{scale invariant} if and only if for any couple of random vectors $\bX$ and $\bY$
    \begin{equations}
    \label{SI}
    \delta(Q_1\bX,Q_2\bY)=\delta(\bX,\bY)
    \end{equations}
    where $Q_1=diag(\bq_1)$ and $Q_2=diag(\bq_2)$ are two diagonal matrices whose diagonal values are positive, thus $\bq_1$ and $\bq_2$ have positive entries.
    %
\end{definition}
 
For a better understanding of the practical importance of condition \fer{SI}, consider the application of multidimensional inequality measures to quantify the difference between two countries, for example in terms of the distribution of income, wealth and consumption  among their citizens. 
It is important that the discrepancy between the two population remains the same, regardless of whether income is expressed in Dollars, Euro, or another currency as we do not want our notion of discrepancy to be dependent on the currency at hand.
This property corresponds to have a \emph{scaling invariant} discrepancy measure. 

\subsubsection{The Uniform Redistribution Property}
\label{sec:UR}
The uniform redistribution property has been developed for inequality indexes to relate the sparsity of two random vectors that are one the translation of the other.
In economic terms, the uniform redistribution property ensures that if we add the same amount of income to all individuals of nation the wealth inequality decreases.
More formally, an inequality index $\tau:\PP(\erre^n)\to \erre$ satisfies the uniform redistribution property if 
\begin{equation}
    \label{UR_fact}
\tau({\bf X}+ \bC) \leq \tau({\bf X}), 
\end{equation}
for any constant positive vector $\bC$.
When it comes to scale invariant discrepancies, the inequality in \eqref{UR_fact} loses its meaning since adding the same quantity to two random vector that are measured with respect to different unit of measures does not necessarily lead to two probability measures that are more even or similar.
For example, giving one thousand american dollars to the american population has a different effect than giving one thousand yen to the japanese population as these quantities entail a different amount of wealth.
For this reason, in order to define a meaningful generalization to \eqref{UR_fact}, we need to account for the difference induced by the change in unit measure through a suitable whitening process.
%
\begin{definition}[Uniform redistribution property]
\label{def:redprop}
Let $\bX\sim\mu$ and $\bY\sim \nu$ be two random vectors and their associated probability laws.
Given a scale invariant discrepancy $\delta$, we say that $\delta$ satisfies the uniform redistribution with respect to a whitening process $\mu\to W_{\mu}$ if and only if 
\begin{equation}
    \label{UR}
\delta({\bf X}+ \bf{C}_1, {\bf Y} + {\bf C}_2) \le \delta({\bf X}, {\bf Y})+|W_\mu\bf C_1-W_\nu\bf C_2|, 
\end{equation}
for any pair of constant positive vector $\bf C_1$ and $\bf C_2$ with non negative components.
\end{definition}
%
The role of the term $|W_\mu\bf C_1-W_\nu\bf C_2|$ in \eqref{UR} is to quantify how much the different measure unit of the constants $\bC_1$ and $\bC_2$ affect the distribution that are described by $\bX$ and $\bY$.
In particular, we notice that, if $W_\mu \bC_1 = W_\nu \bC_2$, it means that adding $\bC_1$ to $\bX$ is the same as adding $\bC_2$ to $\bY$, hence the term $|W_\mu\bC_1-W_\nu\bC_2|$ vanishes and we recover
\[
\delta(\bX+\bC_1,\bY+\bC_2)\le \delta(\bX,\bY).
\]
Moreover, if $\bC=\bC_1=\bC_2$, then we have that $W_\mu\bC=W_\nu\bC$ for every $\bC\in\erre^n$ if and only if $W_\mu=W_\nu$, i.e. if and only if $\bX$ and $\bY$ are described through the same unit measure.
%



\section{Multivariate scaling-invariant discrepancies }
\label{sec:3}
In this section, we apply the notions on whitening to introduce and study three novel discrepancies: the White Wasserstein discrepancy, the White Fourier discrepancy, and the Gini discrepancy. 
The key idea behind all three discrepancies is to pre-process the probability measures via a suitable whitening to recover the scale invariance.
%

%

\subsection{The White Wasserstein discrepancy}
\label{sec:transport}
In this section, we introduce and study a discrepancy based on the Wasserstein Distance.
First of all, we notice that $W_1$ defined in \eqref{wdp} is not scale invariant, however, by suitably choosing a whitening processes, it is possible to recover a scale invariant Wasserstein discrepancy.
%

%
%
In agreement with Section \ref{sec:SI}, we consider the ZCA-cor whitening processes, thus given a $\bX\sim\mu$ random vector, we denote with $W^{ZCA}_\mu$ the ZCA-cor whitening process associated with $\bX$.
Given two probability measure $\mu$ and $\nu$ in $\mathcal{P}_2(\R^n)$, the idea is to use the Wasserstein distance to measure the discrepancy between the two whitened probability measures.
We then define
\begin{align}
    \label{gopt}
    & \WWD(\mu, \nu) = \inf_{\pi \in \Pi(\mu^*,\nu^*)} \, \int_{\R^n\times\R^n} |\bx^* - \by^*| \, \pi(d\bx^*,d\by^*),  \\
    \label{gd2}
    &\bx^* =  W^{ZCA}_\mu \bx, \qquad \by^* =W^{ZCA}_\nu \by,
\end{align}
where $\mu^*$ and $\nu^*$ are the probability measures obtained by whitening $\mu$ and $\nu$, respectively.
The value in \eqref{gopt} is the \emph{White Wasserstein discrepancy} between $\mu$ and $\nu$.
Owing to the scale-stability of the ZCA correlation whitening process, we infer that \eqref{gopt} is scale invariant.
Moreover, due to the sub-additivity with respect to convolution of $W_1$, we have that \eqref{gopt} possesses the uniform redistribution property as well.

\begin{theorem}
\label{thm:WWD_prop}
    The discrepancy $\WWD$, defined in \eqref{gopt}, is scale invariant and satisfies the uniform redistribution property.
\end{theorem}

\begin{proof}
    The scale invariance of $\WWD$ follows from the scale stability of the ZCA-correlation whitening process.
    Indeed, let $Q$ be a diagonal matrix whose elements on the diagonal are positive.
    Then we have that
    \[
        W_\mu^{ZCA}\bX=W_{\mu^{Q}}^{ZCA}(Q\bX),
    \]
    hence $\WWD$ is scale invariant.
    We now consider the uniform redistribution property.
    Let $\bX\sim\mu$ and $\bY\sim\nu$ be two random vectors.
    Given $\bC_1$ and $\bC_2$ two positive constant vectors, we have that the ZCA-correlation whitening matrix associated with $\bX$ and $\bX+\bC_1$ are the same.
    The same holds for $\bY$ and $\bY+\bC_2$.
    We then have
    \begin{align*}
    W_1(\bX+&\bC_1,\bY +\bC_2)=
    W_1(W_\mu^{ZCA} (\bX+\bC_1), W_\nu^{ZCA} (\bY+\bC_2))\\
    &\leq W_1(W_\mu^{ZCA} \bX, W_\nu^{ZCA} \bY)+ W_1(W_\mu^{ZCA} \bC_1, W_\nu^{ZCA} \bC_2).
\end{align*}
Since $\WWD(\mu,\nu)=W_1(W_\mu^{ZCA} \bX, W_\nu^{ZCA} \bY)$ and $W_1(W_\mu^{ZCA} \bC_1, W_\nu^{ZCA} \bC_2)=|W_\mu^{ZCA} \bC_1 - W_\nu^{ZCA} \bC_2|$, we conclude the proof.
\end{proof}

It is also worthy of notice that the whitening process affects the metric properties of the Wasserstein Distance.
Indeed, we have that $\WWD(\mu,\nu)=0$ does not imply $\mu=\nu$.
For example, $\WWD(\mu,\nu)=0$ whenever $\mu$ is the probability distribution associated to the random vector $\bX$ and $\nu$ is the probability distribution associated to $\alpha\bX$, for any $\alpha$ positive constant.
To conclude, we show that the $\WWD$ between two Gaussian distributions can be computed explicitly.
%

\begin{proposition}
    Let $\mu \sim \mathcal{N}(\bm_1,\Sigma_1)$, and $\nu \sim \mathcal{N}(\bm_2,\Sigma_2)$ be two Gaussian distributions. 
    Then, 
    \begin{equation}
        \label{div-gau2}
        \WWD(\mu,\nu) =  \left|W_\mu^{ZCA} \bm_1- W_\nu^{ZCA} \bm_2 \right|.
    \end{equation}
\end{proposition}

\begin{proof}
    Performing the change of variable in \eqref{gd2}, we can identify all plans $\pi \in \Pi(\mu,\nu)$ with plans in $\pi^* \in \Pi(\mu^*,\nu^*)$, via the mapping 
    \[
    \pi \to \pi^* = (\bx^*,\by^*)_\# \pi,
    \]
    where $\mu^*$ and $\nu^*$ are the probability distributions associated with $\bX^*$ and $\bY^*$, respectively.
    It is easy to check that $\mu^* \sim \mathcal{N}(\bm_1^*,Id),$ and $\nu^* \sim \mathcal{N}(\bm_2^*,Id),$ with $\bm_1^* = W_\mu^{ZCA} \bm_1$ and $\bm_2^* = W_\nu^{ZCA} \bm_2$, respectively. 
    Then the same change of variable applied to the integral in \eqref{gopt} yields
    \[
    \WWD(\mu,\nu) = W_1(\mu^*,\nu^*),
    \]
    where $W_1$ is the classical $1$-Wasserstein distance of \eqref{wdp}. 
    By Jensen's inequality, we have 
    \[
    \left|W_\mu^{ZCA} \bm_1- W_\nu^{ZCA} \bm_2 \right| \leq W_1(\mu^*,\nu^*) \leq W_2(\mu^*,\nu^*) = \left|W_\mu^{ZCA} \bm_1- W_\nu^{ZCA} \bm_2 \right|,
    \]
    via formula \eqref{otg2}, applied to \emph{white Gaussians}.
\end{proof}

\begin{remark}
    Lastly, notice that through an argument similar to the one used for the Wasserstein Distance, we can define a Whitened Fourier Based Metric by setting
    \[
        WF(\mu,\nu)=d_1(\mu^*,\nu^*),
    \]
    where $\mu^*$ and $\nu^*$ are the probability distributions associated with $W_\mu^{ZCA}\bX$ and $W_\nu^{ZCA}\bY$, respectively, while $d_1$ is the Fourier Based Metric defined in \eqref{fm1}.
    Following the same argument used for the $\WWD$, we have that $WF$ is scale invariant and possesses the uniform redistribution property.
\end{remark}

\subsection{The White Fourier discrepancy}
\label{sec:Fourier}

In a recent paper, the possibility to make use of Fourier metrics to measure the concentration of measures has been outlined.\cite{To1} 
There, a description of one-dimensional  classical inequality indices, like Gini and Pietra, in terms of the Fourier transform  led to the introduction of a new index, purely based on the Fourier transform, which revealed very useful to compute concentration in case of probability densities expressible only in terms of the Fourier transform (stable laws, Poisson distribution, and others).\cite{giudici2024measuring}
%
%
As in Section \ref{sec:transport}, let $\mu,\nu \in \mathcal{P}_2(\R^n)$, and let $W_\mu^{ZCA}$ and $W_\nu^{ZCA}$ be the ZCA-cor whitening process associated with $\mu$ and $\nu$, respectively.
We can then express the Fourier transform of the whitened probability measures $\mu^*$ and $\nu^*$ as it follows
\begin{equation}
    \label{four*}
  \int_{\R^n} e^{-i \bxi^T \bx^*} \, \mu^*(d\bx^*) =  \int_{\R^n} e^{-i \bxi^T W^{ZCA}_\mu \bx} \, \mu(d\bx) =\int_{\R^n} e^{-i (W^{ZCA}_\mu\bxi)^T\bx} \, \mu(d\bx) = \fo_1(\bxi^*),
\end{equation}
where $\hat f_1$ is the Fourier transform of $\mu$ and $\bxi^* = W^{ZCA}_\mu\bxi$.
Analogously
\[
\int_{\R^n} e^{-i \bieta^T \by^*} \, \nu^*(d\by^*) =\fo_2(\bieta^*), \quad \bieta^* = W^{ZCA}_\nu\bieta,
\]
where $\hat f_2$ is the Fourier transform of $\nu$ and $\bieta^* = W^{ZCA}_\nu\bieta$.
%
%
%
Given $\mu,\nu \in \mathcal{P}_2(\R^n)$, we can then define a discrepancy between $\mu$ and $\nu$ as
\begin{equation}
    \label{div-fou}
\WFD(\mu,\nu) =\sup_{\bxi\in \R^n}\left| \fo_1^*(\bxi) \nabla \fo^*_2(\bxi) -\fo_2^*(\bxi) \nabla \fo^*_1(\bxi)\right|.
\end{equation}
First, we show that the discrepancy defined in \eqref{div-fou} possess the properties we are interested in.

\begin{theorem}
    The Fourier Whitened Discrepancy is Scale Invariant and satisfies the uniform redistribution property.
\end{theorem}

\begin{proof}
Following the same argument used to prove Theorem \ref{thm:WWD_prop}, the scale invariancy follows by the scale stability of the ZCA-correlation whitening process.
Let us consider the uniform redistribution property.
Let $\bX\sim\mu$ and $\bY\sim\nu$ be two random vectors and let $\bC_1$ and $\bC_2$ be two constant vectors whose components are non-negative.
Let us denote with $\hat f_1^*$ and $\hat f_2^*$ the Fourier transform of the whitened vectors $\bX^*$ and $\bY^*$, respectively.
Owing to the properties of Fourier transform, we have that
\begin{align}
    \hat f_{1,\bC_1}^*(\bxi)&=\hat f_{1}^*(\bxi)e^{-i\bxi W_\mu^{ZCA}\bC_1},\\
    \nabla \hat f_{1,\bC_1}^*(\bxi)&=\nabla\hat f_{1}^*(\bxi)e^{-i\bxi W_\mu^{ZCA}\bC_1}-iW_\mu^{ZCA}\bC_1\hat f_{1}^*(\bxi)e^{-i\bxi W_\mu^{ZCA}\bC_1},
\end{align}
where $\hat f_{1,\bC_1}$ is the Fourier transform of $(\bX+\bC_1)^*$.
Similarly, we have that
\begin{align}
    \hat f_{2,\bC_2}^*(\bxi)&=\hat f_{2}^*(\bxi)e^{-i\bxi W_\nu^{ZCA}\bC_2},\\
    \nabla \hat f_{2,\bC_2}^*(\bxi)&=\nabla\hat f_{2}^*(\bxi)e^{-i\bxi W_\nu^{ZCA}\bC_2}-iW_\nu^{ZCA}\bC_2\hat f_{2}^*(\bxi)e^{-i\bxi W_\nu^{ZCA}\bC_2},
\end{align}
where $\hat f_{2,\bC_2}$ is the Fourier transform of $(\bY+\bC_2)^*$.
Then, denoted with $\mu_{\bC_1}$ and $\nu_{\bC_2}$ the probability law of $\bX+\bC_1$ and $\bY+\bC_2$, respectively, we have that
\begin{align*}
    \WFD&(\mu_{\bC_1},\nu_{\bC_2})=\sup_{\bxi\in \R^n}\Big| \fo_{1,\bC_1}^*(\bxi) \nabla \fo^*_{2,\bC_2}(\bxi) -\fo_{2,\bC_2}^*(\bxi) \nabla \fo^*_{1,\bC_1}(\bxi)\Big|\\
    &=\sup_{\bxi\in \R^n}\Big|\hat f_{1}^*(\bxi)e^{-i\bxi W_\mu^{ZCA}\bC_1}\Big(\nabla\hat f_{2}^*(\bxi)e^{-i\bxi W_\nu^{ZCA}\bC_2}-iW_\nu^{ZCA}\bC_2\hat f_{2}^*(\bxi)e^{-i\bxi W_\nu^{ZCA}\bC_2}\Big)\\
    &\quad-\hat f_{2}^*(\bxi)e^{-i\bxi W_\nu^{ZCA}\bC_2}\Big(\nabla\hat f_{1}^*(\bxi)e^{-i\bxi W_\mu^{ZCA}\bC_1}-iW_\mu^{ZCA}\bC_1\hat f_{1}^*(\bxi)e^{-i\bxi W_\mu^{ZCA}\bC_1}\Big)\Big|\\
    &\le \sup_{\bxi\in \R^n}\Big|\hat f_{1}^*(\bxi)\nabla\hat f_{2}^*(\bxi)e^{-i\bxi (W_\mu^{ZCA}\bC_1+W_\nu^{ZCA}\bC_2)}-\hat f_{2}^*(\bxi)\nabla\hat f_{1}^*(\bxi)e^{-i\bxi (W_\mu^{ZCA}\bC_1+W_\nu^{ZCA}\bC_2)}\Big|\\
    &\quad+\sup_{\bxi\in \R^n}\Big|\Big(W_\mu^{ZCA}\bC_1-W_\nu^{ZCA}\bC_2\Big)\hat f_{1}^*(\bxi)\hat f_{2}^*(\bxi)e^{-i\bxi (W_\mu^{ZCA}\bC_1+W_\nu^{ZCA}\bC_2)}\Big|\\
    &\le \sup_{\bxi\in \R^n}\Big|\hat f_{1}^*(\bxi)\nabla\hat f_{2}^*(\bxi)-\hat f_{2}^*(\bxi)\nabla\hat f_{1}^*(\bxi)\Big|+\sup_{\bxi\in \R^n}\Big|\Big(W_\mu^{ZCA}\bC_1-W_\nu^{ZCA}\bC_2\Big)\hat f_{1}^*(\bxi)\hat f_{2}^*(\bxi)\Big|\\
    &\le \sup_{\bxi\in \R^n}\Big|\hat f_{1}^*(\bxi)\nabla\hat f_{2}^*(\bxi)-\hat f_{2}^*(\bxi)\nabla\hat f_{1}^*(\bxi)\Big|+\sup_{\bxi\in \R^n}\Big|W_\mu^{ZCA}\bC_1-W_\nu^{ZCA}\bC_2\Big|\\
    &=\WFD(\mu,\nu)+\Big|W_\mu^{ZCA}\bC_1-W_\nu^{ZCA}\bC_2\Big|,
\end{align*}
which concludes the proof.
\end{proof}

We now derive the Fourier discrepancy between two Gaussian distributions. 
Let $\bX_1\sim\mu = \mathcal{N}(\bm_1,\Sigma_1)$ and $\bX_2\sim\nu = \mathcal{N}(\bm_2,\Sigma_2)$ two multidimensional Gaussian distributions, such that $\bm_i >0$, $i=1,2$.
Since for $i =1,2$ we have that
\[
\fo^*_i(\bxi) = \fo_i(\bxi^*) = \exp\left\{ -i (\bxi^*)^T\, \bm_i - \frac 12 |\bxi|^2\right\},
\]
one infer that
\[
\nabla \fo^*_i(\bxi) = -(i \bm_i^* + \bxi)\exp\left\{ -i (\bxi^*)^T\, \bm_i - \frac 12 |\bxi|^2\right\}, 
\]
where $\bm_i^*$ is the expected value of the respective whitened random vector $\bX^*_i$. 
Hence
\[
\WFD(\mu,\nu) =\sup_{\bxi\in \R^n}\left| \bm_1^* -\bm_2^* \right|\, e^{-|\bxi|^2}=\left| \bm_1^* -\bm_2^* \right|.
\]
Thus, the Fourier-based discrepancy between two Gaussian densities is given by the expression
\begin{equation}
    \label{div-gau}
\WFD(\mu,\nu) =\left| W_\mu^{ZCA} \bm_1- W_\nu^{ZCA} \bm_2 \right|.
\end{equation}

\begin{remark}
    Notice that, on the class of Gaussian distributions the White Fourier discrepancy and the White Wasserstein discrepancy attain the same value.
\end{remark}

\subsubsection*{The Relation with the Fourier sparsity index.}

This definition of the Fourier Whitened Discrepancy is consistent with the definition of multidimensional and scaling invariant inequality index introduced in Ref. \cite{giudici2024measuring}. 
Indeed, by setting $\fo_2^* (\bxi) = \exp\left\{-i\bxi^T \bm_1^* \right\}$,\footnote{i.e. by choosing $\nu$ as a Dirac delta function concentrated at the mean value of $\mu$, that is $\bm_1^*$}  
and dividing \fer{div-fou} by $2|\bm_1^*|=\mathbb{E}(\bX)+\mathbb{E}(\bm_1^*)$, we obtain
\begin{equation}\label{ine-T*}
\tau(\mu) =
 \sup_{\bxi \in \R^n} \frac{ \left| \nabla \fo^*_1(\bxi = {\bf 0}) \fo^*_1(\bxi) - \nabla \fo^*_1(\bxi) \right|}{2\,| \nabla \fo^*_1(\bxi = {\bf 0})|},
\end{equation}
which is the inequality index $\tau(\mu)$ introduced in Ref. \cite{giudici2024measuring}.
Notice that the scaling invariant inequality index $\tau$ of a multivariate Gaussian distribution $\mu \sim \mathcal{N}(\bm,\Sigma)$ takes the form 
\begin{equation}
    \label{T-Gauss}
\tau(\mu) = \frac1{2\sqrt e} \frac 1{\sqrt{\bm^T\Sigma^{-1}\bm}},
\end{equation}
a function of the Mahalanobis distance from the origin.\cite{giudici2024measuring}
The inequality index in \fer{T-Gauss} is proportional to the multivariate coefficient of variation considered by Voinov and Nikulin in their book,\cite{VN} which, is defined as
\begin{equation}
    \label{voi}
C_{VN}(\mu) =  \frac 1{\sqrt{\bm^T\Sigma^{-1}\bm}}.
\end{equation}
Among all proposals of multivariate coefficient of variations considered in the literature, $C_{VN}$ is the only one that possess the scaling invariance property.\cite{Aer}
%
%
Moreover, it is interesting to remark that the quantity defined in \fer{div-gau} could be employed to define a coefficient of variation for any pair of measures in $\mathcal P_2(\R^n)$.
%
%
The advantage of working with an explicit expression which depends only on moments of the first two orders, and is build to possess the scaling invariant property is evident.

\subsection{The Gini Discrepancy}

Finally, we introduce a discrepancy inspired by the multivariate Gini Index considered in \cite{AGT}.

\begin{definition}[Gini discrepancy]
Given $\mu,\nu\in\PP(\R^n)$, we define the Gini discrepancy as it follows
\begin{equation}
    \label{eq:refMDformula}
    \WGD(\mu,\nu)=\int_{\R^n\times \R^n}|W^{ZCA}_\mu \bx-W^{ZCA}_\nu \by|\mu(d\bx)\mu(d\by),
\end{equation}
where $W_\mu^{ZCA}$ and $W_\nu^{ZCA}$ are the whitening matrix associated with $\mu$ and $\nu$, respectively.
\end{definition}

Owing to the scale stability of the ZCA-correlation whitening process, the Gini discrepancy is scale invariant as well.

\begin{theorem}
    The Gini Discrepancy is scale invariant and possess the uniform redistribution property.
\end{theorem}

\begin{proof}
    Following the same argument used to prove Theorem \ref{thm:WWD_prop}, the scale invariancy follows by the scale stability of the ZCA-correlation whitening process.
    To prove that $\WGD$ possess the uniform redistribution property, let $\bX\sim\mu$ and $\bY\sim\nu$ bet two random vectors and let $\bC_1$ and $\bC_2$ be two positive vectors.
    Since the covariance matrix of $\bX$ and $\bX+\bC_1$ are the same and, likewise, the covariance matrix of $\bY$ and $\bY+\bC_2$ are the same, we have that
    \begin{align*}
        \WGD(\bX&+\bC_1,\bY+\bC_2) =\int_{\R^n\times \R^n}|W^{ZCA}_\mu (\bx+\bC_1)-W^{ZCA}_\nu (\by+\bC_2)|\mu(d\bx)\mu(d\by)\\
        &\le \int_{\R^n\times \R^n}|W^{ZCA}_\mu \bx-W^{ZCA}_\nu \by|\mu(d\bx)\mu(d\by)+|W^{ZCA}_\mu \bC_1-W^{ZCA}_\nu\bC_2|,
    \end{align*}
    which concludes the proof.    
\end{proof}

\begin{remark}
    Notice that the value of the Gini discrepancy is equal to the cost of selecting the independent transportation plan between $\mu$ and $\nu$ in \eqref{gopt}.
\end{remark}


We now derive the Gini discrepancy between two Gaussian Distributions.
Let $\bX\sim\mu=\mathcal{N}(\bm_1,\Sigma_1)$ and $\bY\sim\nu=\mathcal{N}(\bm_2,\Sigma_2)$ be two Gaussian distribution.
We then have that
\begin{equation}
    \label{eq:DGgaussian}
    \WGD(\mu,\nu)=\int_{\R^n\times\R^n}|\bx^*-\by^*|\mu^*(d\bx^*)\nu^*(d\by^*),
\end{equation}
where $\mu^*$ and $\nu^*$ are the Gaussian distribution associated with the whitened random vectors $\bX^*$ and $\bY^*$ respectively. 
Since both $\bX^*$ and $\bY^*$ are Gaussian random vectors, then $\bX^*-\bY^*$ is a Gaussian random vector whose components are independent, whose mean is $\bm^*_1-\bm^*_2$, and whose covariance matrix is $\sqrt{2}Id$.
%
Although the quantity in \eqref{eq:DGgaussian} does not have an analytic formula, we provide an easy-to-compute upper bound.
%

\begin{proposition}
    Let $\mu \sim \mathcal{N}(\bm_1,\Sigma_1)$ and $\nu \sim \mathcal{N}(\bm_2,\Sigma_2)$.
    Then, we have that
    \begin{equation}
        \label{goptg}
        \WGD(\mu,\nu) \le \sqrt{2n+|\bm_1^*-\bm_2^*|^2}.
    \end{equation}
\end{proposition}

\begin{proof}
    By definition, we have that
    \begin{align*}
        \WGD(\mu,\nu)&=\int_{\R^n\times\R^n}|\bx^*-\by^*|\mu^*(d\bx^*) \nu^*(d\by^*) \le \sqrt{\int_{\R^n\times\R^n}|\bx^*-\by^*|^2 \mu^*(d\bx^*) \nu^*(d\by^*)}\\
         &\le\sqrt{\int_{\R^n\times\R^n}\sum_{i=1}^n|x_i^*-y_i^*|^2\mu^*(d\bx^*) \nu^*(d\by^*)}\\
         &=\sqrt{\sum_{i=1}^n\int_{\R^n\times\R^n}|x_i^*-y_i^*|^2\mu^*(d\bx^*) \nu^*(d\by^*)}\\
         &=\sqrt{\sum_{i=1}^n\int_{\R^n\times\R^n}\big((x_i^*)^2+(y_i^*)^2-2x_i^*y_i^*\big)\mu^*(d\bx^*) \nu^*(d\by^*)}.
    \end{align*}
    It is easy to see that
    \[
        \int_{\R^n\times\R^n}\big((x_i^*)^2+(y_i^*)^2-2x_i^*y_i^*\big)\mu^*(d\bx^*) \nu^*(d\by^*)=2+\big((m_1)_i^*-(m_2)^*_i\big)^2.
    \]
    Indeed, since $\bX^*$ is a whitened random vector, we have that
    \[
        \int_{\R^n\times\R^n}(x_i^*)^2 \mu^*(d\bx^*) =\int_{\R^n\times\R^n}(x_i^*-m_1^*)^2\mu^*(d\bx^*)+(m_1^*)^2_i=1+(m_1^*)^2_i.
    \]
    Likewise
    \[
        \int_{\R^n\times\R^n}(y_i^*)^2\nu^*(d\by^*)=1+(m_2^*)^2_i.
    \]
    Finally, we have that
    \[
        2\int_{\R^n\times\R^n}x_i^*y_i^*\mu^*(d\bx^*) \nu^*(d\by^*)=2\Big(\int_{\R^n}x^*_i\mu^*(d\bx^*) \Big)\Big(\int_{\R^n}y_i^*\nu^*(d\by^*)\Big)=2(m_1^*)_i(m_2^*)_i.
    \]
    Putting everything together, we retrieve equation \eqref{goptg}, concluding the proof.
\end{proof}

\begin{remark}[Basic estimates and equivalence with other distances]
By Jensen's inequality, it is easy to see that, for all $\pi \in \Pi(\mu,\nu)$, we have 
\begin{equation*}
    \WWD(\mu,\nu) \geq |\bm_1^* - \bm_2^*|, \qquad \bm_1^* =  W^{ZCA}_\mu \bm_1, \quad \bm_2^* =  W^{ZCA}_\nu \bm_2.
\end{equation*}
This bound is sharp and the equality is attained when both $\mu$ and $\nu$ are Gaussian. 
Owing again to the Jensen's inequality, we have that
\[
\WWD(\mu,\nu) \leq W^{1/2}_2(\mu_{1}^*,\mu^*_{2}).
\]
Under suitable conditions on $\mu_{2}^*$, the right-hand side of the last display can be further bounded from above by the Kullback-Leidler discrepancy $\mathrm{KL}(\mu^{*},\nu^{*})$, and finally by the \emph{relative Fisher information} between $\mu^{*}$ and $\nu^{*}$.\cite{otto2000generalization} 
Recall finally that $KL(\mu,\nu)$ equals to the Mahalanobis distance between $\mu$ and $\nu$, when $\mu$ and $\nu$ are Gaussian distributions with the same covariance matrix.\cite[Section 2.1]{barhen1995generalization}
\end{remark}


\section{The Connections between the Three discrepancies}

In this section, we relate the three discrepancies we have introduced in Section \ref{sec:3}.
First, we relate the White Fourier discrepancy and the Gini discrepancy.

\begin{proposition}\label{prop41}
    Given two probability measures $\mu,\nu$, we have that 
    \begin{align*}
        \WGD(\mu,\nu) &\leq \WFD(\mu,\nu) + \int_{\R^n}|\bx^*-\bm_1^*|\mu^*(d\bx^*) + \int_{\R^n}|\by^*-\bm_2^*|\nu^*(d\by^*), \\
        \WFD(\mu,\nu) &\leq \WGD(\mu,\nu), 
    \end{align*}
    where $\bm_1^*$ is the mean of $\mu^*$ and $\bm_2^*$ is the mean of $\nu^*$.
\end{proposition}

\begin{proof}
    First, we show that 
    \[
        \WGD(\mu,\nu) \leq \WFD(\mu,\nu) + \int_{\R^n}|\bx^*-\bm_1^*|\mu^*(d\bx^*) + \int_{\R^n}|\by^*-\bm_2^*|\nu^*(d\by^*).
    \]
    By definition, we have that
    \begin{align*}
     \WGD(\mu,\nu) &= \int_{\R^n}\int_{\R^n}|\bx^*-\by^*| \mu^*(d\bx^*) \nu^*(d\by^*) \\
        &\leq  \int_{\R^n}\int_{\R^n}|\bx^*-\bm_1^*+\bm_1^*-\bm_2^*+\bm_2^*-\by^*|\mu^*(d\bx^*) \nu^*(d\by^*)\\
        &\leq \int_{\R^n}\int_{\R^n}\Big(|\bx^*-\bm_1^*|+|\bm_1^*-\bm_2^*|+|\bm_2^*-\by^*|\Big)\mu^*(d\bx^*) \nu^*(d\by^*)\\
        &\leq  \Bigg(\int_{\R^n}\int_{\R^n}\Big(|\bx^*-\bm_1^*|+|\bm_2^*-\by^*|\Big)\mu^*(d\bx^*) \nu^*(d\by^*)+|\bm_1^*-\bm_2^*|\Bigg)
    \end{align*}
    Denoted with $\hat f_1$ and $\hat f_2$ the Fourier transforms of $\mu$ and $\nu$ respectively, we observe that $\bm_1^*=i\nabla \hat f(0)$, $\bm_2^*=i\nabla \hat g(0)$, and $\hat f(0)=\hat g (0)=1$, thus
    \[
        \bm_1^*-\bm_2^*=-i(\nabla \hat f(0)\hat g(0)-\nabla \hat g(0)\hat f(0)).
    \]
    We then conclude that
    \begin{align}
        |\bm_1^*-\bm_2^*|&=|\nabla \hat f(0)\hat g(0)-\nabla \hat g(0)\hat f(0)|\\
        &\leq \sup_{\bxi\in\R^n}|\nabla \hat f(\bxi)\hat g(\bxi)-\nabla \hat g(\bxi)\hat f(\bxi))|,
    \end{align}
    which allows us to conclude the first part of the proof.
    To conclude, we show that
    \[
        \WFD(\mu,\nu) \leq \WGD(\mu,\nu).
    \]
    Indeed, let $\mu^*\otimes \nu^*$ be the independent transportation plan between $\mu^*$ and $\nu^*$, then the Fourier transform of $\mu^*\otimes \nu^*$ is $\hat f\hat g$, where $\hat f$ is the Fourier transform of $\mu^*$ and $\hat g$ is the Fourier transform of $\nu^*$.
    In particular, we have that
    \[
    (\nabla\hat f(\bxi))\hat g(\bieta)=-i\int_{\R^n}\int_{\R^n}\bx^* e^{-i\bx^*\bxi-i\by^*\bieta}\mu^*\otimes \nu^*(d\bx^*,d\by^*)
    \]
    and
    \[
    \hat f(\bxi)(\nabla\hat g(\bieta))=-i\int_{\R^n}\int_{\R^n}\by^* e^{-i\bx^*\bxi-i\by^*\bieta}\mu^*\otimes \nu^*(d\bx^*,d\by^*).
    \]
    Putting everything together, we infer
    \begin{align*}
        \sup_{\bxi\in\R^n}\big|\nabla \hat f(\bxi)\hat g(\bxi)-\nabla \hat g(\bxi)\hat f(\bxi))\big|&=\Big|\int_{\R^n}\int_{\R^n}-i(\bx^*-\by^*) e^{-i\bx^*\bxi-i\by^*\bieta}\mu^*\otimes \nu^*(d\bx^*,d\by^*)\Big|\\
        &\le \int_{\R^n}\int_{\R^n}|\bx^*-\by^*|\mu^*\otimes \nu^*(d\bx^*,d\by^*).
    \end{align*}
\end{proof}

We then relate the Gini discrepancy and the White Wasserstein discrepancy.

\begin{proposition}\label{prop42}
    Given two probability measures $\mu$ and $\nu$, we have that 
    \begin{align}
    \label{bound1Gini_WW}
        \WGD(\mu,\nu)&\le \WWD(\mu,\nu)+\min\Big\{G(\mu),G(\nu)\Big\}, \\
    \label{bound2Gini_WW}    
        \WWD(\mu,\nu) &\leq \WGD(\mu,\nu), 
    \end{align}
    where 
    \[
        G(\mu)=\int_{\R^n}\int_{\R^n}|W_\mu^{ZCA}(\bx-\bx')|\mu(d\bx)\mu(d\bx')
    \]
    and $G(\nu)$ is defined similarly.
\end{proposition}

\begin{proof}
    The inequality 
    \[
        \WWD(\mu,\nu) \leq \WGD(\mu,\nu)
    \]
    follows from the definition of $\WWD$, since $\mu^*\otimes\nu^*\in\Pi(\mu^*,\nu^*)$.
    Let us now show that 
    \[
         \WGD(\mu,\nu)\le \WWD(\mu,\nu)+\min\Big\{G(\mu),G(\nu)\Big\}.
    \]
    Let $\pi^*$ be the optimal transportation plan between $\mu^*$ and $\nu^*$.
    Since $\mu^*$ is absolutely continuous, we have that there exists a function $T$ such that $T_\#\mu^*=\nu^*$ and
    \[
        \int_{\R^n}|\bx^*-T(\bx^*)|\mu^*(d\bx^*)=\int_{\R^n\times\R^n}|\bx^*-\by^*|\pi^*(d\bx^*,d\by^*).
    \]
    We then have that
    \begin{align*}
        \WGD(\mu,\nu)&=\int_{\R^n\times\R^n}|\bx^*-\by^*|\mu^*(d\bx^*)\nu^*(d\by^*)\\
        &=\int_{\R^n\times\R^n}|\bx^*-\by^*|\pi^*(d\bx^*,d\by^*)\\
        &\quad+\int_{\R^n\times\R^n}|\bx^*-\by^*|(\mu^*(d\bx^*)\nu^*(d\by^*)-\pi^*(d\bx^*,d\by^*))\\
        &=\WWD(\mu,\nu)+\int_{\R^n\times\R^n}|\bx^*-\by^*|(\mu^*(d\bx^*)\nu^*(d\by^*)-\pi^*(d\bx^*,d\by^*)).
    \end{align*}
    Moreover, we have that
    \begin{align*}
        \int_{\R^n\times\R^n}|\bx^*-\by^*|(&\mu^*(d\bx^*)\nu^*(d\by^*)-\pi^*(d\bx^*,d\by^*))\\
        &=\int_{\R^n\times\R^n}|\bx^*-\by^*|\mu^*(d\bx^*)\nu^*(d\by^*)-\int_{\R^n}|\bx^*-T(\bx^*)|\mu^*(d\bx^*)\\
        &=\int_{\R^n}\Big(\int_{\R^n}|\bx^*-\by^*|\nu^*(d\by^*)-|\bx^*-T(\bx^*)|\Big)\mu^*(d\bx^*)\\
        &=\int_{\R^n}\int_{\R^n}\Big(|\bx^*-\by^*|-|\bx^*-T(\bx^*)|\Big)\nu^*(d\by^*)\mu^*(d\bx^*)\\
        &\le \int_{\R^n}\int_{\R^n}|\by^*-T(\bx^*)|\nu^*(d\by^*)\mu^*(d\bx^*)\\
        &=\int_{\R^n}\int_{\R^n}|\by^*-(\by^*)'|\nu^*(d\by^*)\nu^*(d(\by^*)').
    \end{align*}
    By swapping the roles of $\mu$ and $\nu$, we find
    \[
        \WGD(\mu,\nu)\le \WWD(\mu,\nu)+\int_{\R^n}\int_{\R^n}|\bx^*-(\bx^*)'|\mu^*(d\bx^*)\mu^*(d(\bx^*)'),
    \]
    therefore
    \[
        \WGD(\mu,\nu)\le \WWD(\mu,\nu)+\min\{G(\mu),G(\nu)\}.
    \]
\end{proof}

It is worth noticing that the inequality in \eqref{bound1Gini_WW} is sharp, with equality case corresponding to either $\mu$ or $\nu$ being a Dirac delta.
To conclude, we compare the White Fourier discrepancy with the White Wasserstein discrepancy.
We recall that, given a measure $\gamma$, the non-negative measure $|\gamma|= \gamma_+ + \gamma_-$ is the \emph{total variation} of $\gamma$, according to the Hahn--Jordan decomposition into positive and negative part $\gamma= \gamma_+ - \gamma_-$.

\begin{proposition}\label{prop43}
    Let $\mu,\nu$ be two probability measures.
    Let $\pi^*$ be an optimal transport plan between $\mu_{1}^*$ and $\mu_{2}^*$.
    Then, we have 
    \begin{align}
    \label{bound1}
        \WWD(\mu,\nu) &\leq \WFD(\mu,\nu) + \sqrt{2n - 2 \int_{\R^n\times\R^n} (\bx^*-\bm_{1}^*)^T (\by^*-\bm_{2}^*) \, \pi^*(d\bx^*,d\by^*)} \\
    \label{bound2}    
        \WFD(\mu,\nu) &\leq \WWD(\mu,\nu) + \int_{\R^n\times\R^n} |\bx-\by| \,|\pi^* - \mu^* \otimes \nu^*|(d\bx^*,d\by^*)
    \end{align}
\end{proposition}

\begin{proof}
    Let us start with \eqref{bound1}. Similarly to the proof of Proposition \ref{prop41}, we can bound 
    \begin{align*}
        & \WWD(\mu,\nu) 
        = \int_{\R^n\times\R^n} |\bx^*-\by^*| \, \pi^*(d\bx^*,d\by^*) \\ 
        &\leq |\bm_1^* - \bm_2^*| + \int_{\R^n\times\R^n} |\bx^*-\bm_1^* - \by^* + \bm_2^*| \pi^*(d\bx^*,d\by^*) \\ 
        &\leq  \WFD(\mu,\nu) + \sqrt{ \int_{\R^n\times\R^n} |\bx^*-\bm_1^* - (\by^* - \bm_2^*)|^2 \pi^*(d\bx^*,d\by^*)} \\ 
       & \leq \WFD(\mu,\nu) + \sqrt{ \int_{\R^n\times\R^n} \big(2n - 2(\bx^*-\bm_1^*)^T (\by^*-\bm_{2}^*)\big) \, \pi^*(d\bx^*,d\by^*)}
    \end{align*}
    where $\pi^*$ is any $W_1$-optimal transport plan between $\mu^*$ and $\nu^*$, and the last inequality is implied by $\pi^*$ having whitened marginals.
    For \eqref{bound2}, let us rewrite
    \[
     \WFD(\mu,\nu) = \sup_{\bxi \in \R^n} \left| \int_{\erre^n\times\erre^n} (\bx^*-\by^*)\mathrm{e}^{-i\bxi(\bx^*+\by^*) } \mu^*(d\bx^*) \nu^*(d\by^*) \right|,
    \]
    which is bounded above by 
    \[
    \begin{aligned}
    &\sup_{\bxi \in \R^n} \left| \int_{\erre^n\times\erre^n} \mathrm{e}^{-i\bxi(\bx^*+\by^*)} \, (\bx^*-\by^*) \pi^*(d\bx^*,d\by^*) \right| \\  &\qquad + \sup_{\bxi \in \R^n} \left| \int_{\erre^n\times\erre^n} \mathrm{e}^{-i\bxi(\bx^*+\by^*)} \, (\bx^*-\by^*) (\pi^*(d\bx^*,d\by^*) - \mu^*(d\bx^*)\nu^*(d\by^*)) \right| \\ &\leq  \WWD(\mu,\nu) + \int_{\erre^n\times\erre^n} |\bx^*-\by^*| \big( |\pi^* - \mu^*\otimes\nu^*|\big)(d\bx^*,d\by^*), 
    \end{aligned}
    \]
    which concludes the proof.
    %
\end{proof}

When the measure $\nu^*$ is concentrated on one point, we simplify the bounds of Proposition \ref{prop43} as follows.

\begin{corollary}
    In the framework of Proposition \ref{prop43}, assume that $\nu^* = \delta_{\bm_2^*}$. Then, for all measures $\mu$, we have
    \begin{align*}
        \left|\bm_1^*-\bm_2^*\right| &\leq \WFD(\mu,\nu) \leq \WWD(\mu,\nu) \\
        &= \int_{\erre^n} |\bx^*-\bm_2^*| \, \mu^*(d\bx^*) \leq |\bm_1^*-\bm_2^*| + \int_{\erre^n} |\bx^*-\bm_1^*| \, \mu^*(d\bx^*).
    \end{align*}
\end{corollary}

The two extra terms in the right hand sides of \eqref{bound1}-\eqref{bound2}  capture the sparsity of the optimal plan $\pi^*$, by telling how different it is from $\mu^*_1 \otimes \mu^*_2$.
However, even if these terms \emph{a priori} depend on the plan $\pi^*$ -- which does not admit a closed expression in terms of $\mu$ and $\nu$, but is computable in an efficient way,\cite{cuturi2013sinkhorn} -- those can be easily estimated from above with (suboptimal) quantities depending only on the first and second-order moments of $\mu$ and $\nu$.
In addition, these two terms vanish when $\mu = \nu$, and are uniformly small when the two measures are close-by. 
Then, a trade-off is quantitatively established between $\WWD$ and $\WFD$:
\begin{itemize}
    \item the indicator $\WFD$ is elementary to compute, but very \emph{sparse}, as it does not take into account the coupling between the measures $\mu$ and $\nu$, treating them as independent;
    \item the discrepancy $\WWD$ needs one more step (i.e.~finding an optimal transport plan) to be computed. On the other side, the resulting optimal plan is very concentrated (either on a graph or on a few pairs $(x_i,y_i)_i$ if $\mu$ and $\nu$ are discrete) as discussed in Section \ref{ss:OT}.
    This yields both computational and structural advantages, being \emph{de facto} a dimensional reduction.
\end{itemize}


\section{Application}
\label{sec:examples}

We consider an important economic problem: the impact of Sustainability on the economy and, specifically, the impact of Environmental, Social, and Governance (ESG) factors on company development. 
This is in line with the general aim to improve sustainable policies that are also financially viable. 
Indeed, there is a growing demand for companies to deliver strong financial results while contributing positively to investments' sustainability and ethical impact. 
Thus, investigating the relationship between financial performance and ESG scores as well as understanding whether sustainable factors affect financial performance is a topic of great interest.
To this aim, we consider annual balance sheet data from Small and Medium Enterprises in various sectors across Italy, covering the period from $2020$ to $2022$.
The data source is the Modefinance database, a FinTech company accredited as a Credit Rating Agency by the European Securities and Markets Authority. 
This dataset includes the ESG scores derived from indicators of environmental sustainability, social responsibility, and governance practices, along with essential financial metrics such as revenue, profit, assets, and liabilities.  
We classify the economic sectors into five broader categories, following the Global Industry Classification Standard (GICS), as shown in the following table.

\begin{table}[!ht]
    \centering
    \begin{adjustbox}{width=0.9\textwidth}
    
    \footnotesize
    \begin{tabular}{llll}
        \toprule
        Sector & Detail & Frequency & Percentage \\ 
\hline
        Consumer & Consumer Staples \& Consumer Discretionary & 351 & 33.05 \\ 
        Financials & Banking, Insurance \& Financial Services, and Real Estate & 14 & 1.32 \\ 
        Health.Util & Healthcare \& Essential Utility Services & 55 & 5.18 \\  
        Manufacturing & Materials \& Industrial Activities & 579 & 54.52 \\
        Tech.Com & Information Technology \& Communication Industries & 63 & 5.93 \\ 
        \hline
        Total & & 1062 & 100.00 \\
        \hline
    \end{tabular}
    \end{adjustbox}
    \caption{Distribution of Companies by Sector Classification.}
    \label{Sector-Classification}
\end{table}

After a preliminary cleaning, the dataset under study consists of a total of $1,062$ observations.
In Table \ref{Sector-Classification}, we display the distribution of the Small and Medium Enterprises (SMEs) across sectors. 
Out of a total of $1,062$ companies, $351$ ($33.05\%$) are from the Consumer sector, $14$ ($1.32\%$) from the Financial sector, $55$ ($5.18\%$) from the Health \& Essential Utilities sector, $579$ ($54.52\%$) from the Manufacturing sector, and $63$ ($5.93\%$) from the Technology \& Communication Industries sector.
We now consider a summary analysis of the ESG metrics for the considered companies, which we report in Table \ref{Summary Statistics-ESG}.

\begin{table}[!ht]
    \centering
    \footnotesize
    \begin{tabular}{cllllll}
        \hline
        & Mean & Median & Sdev & Min & Max & Range \\
        \hline
        ESG & 0.65 & 0.66 & 0.11 & 0.28 & 0.93 & 0.65 \\ 
        E.Sc & 0.76 & 0.79 & 0.17 & 0.07 & 0.93 & 0.86 \\ 
        S.Sc & 0.51 & 0.50 & 0.23 & 0.07 & 0.93 & 0.86 \\ 
        G.Sc & 0.62 & 0.64 & 0.15 & 0.21 & 0.93 & 0.71 \\ 
        \hline
    \end{tabular}
    \caption{Summary Statistics for ESG Metrics.}
    \label{Summary Statistics-ESG}
\end{table}

From Table \ref{Summary Statistics-ESG}, we observe that the Overall ESG Scores (ESG) exhibit moderate variability, clustering around a mean of $0.65$, indicating a consistent, though not uniform, performance in sustainability practices.
Environmental Scores (E.Sc) are notably higher on average ($0.76$), suggesting that companies are generally performing better in environmental sustainability. 
In contrast, Social Scores (S.Sc) are considerably lower, averaging around $0.51$, with significant variability, reflecting diverse levels of commitment to social responsibility. 
Governance Scores (G.Sc) demonstrate moderate variability, with an average score of $0.62$, thus indicating that governance practices across companies are relatively consistent but leave room for improvement.
These summary statistics highlight varying levels of commitment to environmental, social, and governance practices among companies. 
Understanding these disparities, and how they impact financial performances, is crucial for policymakers, investors, and other stakeholders who aim to foster sustainable business development. 
To measure the financial performance of the companies, we have extracted three financial indicators from the balance sheets of each company to summarise their economic status and performance.

\begin{table}[!ht]
    \centering
    
    \footnotesize
    \begin{tabular}{cllr}
        \hline
        & Indicator & Code & Description \\ 
        \hline
        1 & Total Asset & TASS & Size of assets. \\ 
    
        2 & Turnover & TOVR &  Size of Sales. \\ 
        
        3 & Shareholders' Funds & SFND & Size of Equity. \\ 
    
        \hline
    \end{tabular}
    
    \caption{Description of Key Financial Metrics.}
    \label{Variable-Definitions}
\end{table}

\begin{table}[!ht]
    \centering
    \footnotesize
    \begin{tabular}{cllllll}
        \hline
        & Mean & Median & Sdev & Min & Max & Range \\
        \hline
        TASS & 173476.54 & 43824.91 & 702804.70 & 1151.23 & 14392422.00 & 14391270.77 \\ 
        SFND & 62341.83 & 15054.90 & 250588.41 & -49091.00 & 5336752.00 & 5385843.00 \\ 
        TOVR & 170644.62 & 43528.31 & 602012.08 & 1288.97 & 10587145.00 & 10585856.03 \\ 
        \hline
    \end{tabular}
    \caption{Summary Statistics for Financial Indicators - 2022 Annual Data.}
    \label{Summary Statistics-FI}
\end{table}

\begin{figure}
	\begin{centering}
		\includegraphics[scale=0.6]{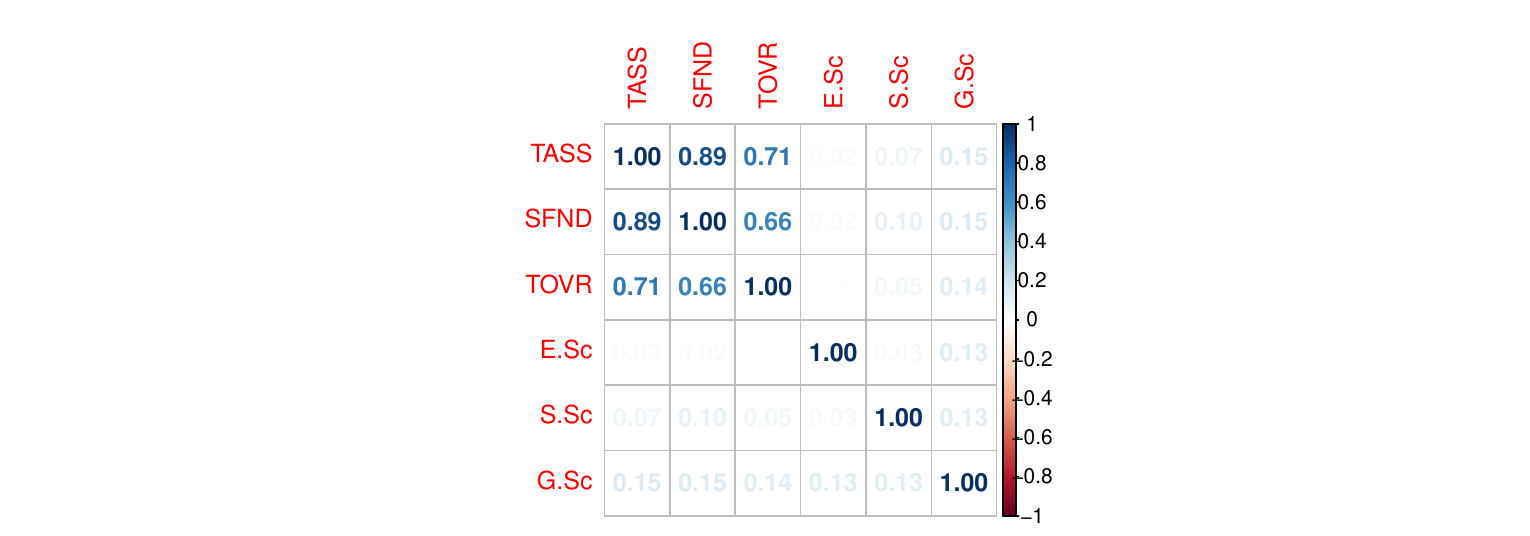}
		\par\end{centering}	
		\caption{Correlation Matrix of Financial Indicators and ESG Scores for 2022.}
  \label{correlations}
\end{figure}

In Table \ref{Variable-Definitions} we present these financial indicators, while in Table \ref{Summary Statistics-FI} we provide a summary statistic of the key financial indicators metrics for $2022$.
The Total Assets (TASS) indicator exhibit a broad range from $1,151.23$ EUR to $14,392,422.00$ EUR, with a high standard deviation indicating significant variability across companies. 
The mean value is relatively high ($173476.54$ EUR), suggesting that there are companies with substantial asset holdings.
Shareholders' Funds (SFND) also show notable variation, with a high standard deviation and some negative values indicating negative equity for certain companies. 
The mean is considerably lower than the maximum, highlighting the impact of a few companies with exceptionally high shareholders' funds.
Turnover (TOVR) displays a large range (from $1,288.97$ EUR to $10,587,145.00$ EUR) and a high standard deviation, reflecting significant differences in revenue generation among companies. 
The mean turnover is elevated by companies with very high revenues.
Overall, the summary statistics of the data reveal significant variability across financial metrics, indicating a wide range of company sizes, financial health, and performance levels, pointing towards high inequality and market concentration.
The correlation matrix in Figure \ref{correlations} presents the relationships between financial indicators and ESG scores for SMEs in $2022$. 

The figure shows that, while Financial indicators are highly correlated with each other, with correlations ranging from $0.66$ to $0.89$, Sustainability indicators are weakly correlated with each other, with a maximum at $0.13$.
Sustainability indicators are also weakly correlated with financial indicators, with maximum values at $0.15$. 
The observed correlations show that it will be quite challenging to build a linear machine learning model that can predict financial variables based on sustainability variables. 
More information will be necessary, such as the sector to which the companies belong.
Alternatively, a non-linear model may be more accurate than a simple linear model.
To address these questions, we now consider three alternative machine learning models to predict financial performance based on ESG scores:
\begin{enumerate*}[label=(\roman*)]
    \item a multivariate regression model, in which the three financial performance variables are explained by the three ESG scores, independently of the sector (LIN);
    \item a similar model, but dependent on sectors (LINS);  
    \item a neural network model with the same variables as the previous model (NNET), and five hidden nodes. 
\end{enumerate*} 
To compare the models, we randomly split the data into a $80\%$ training sample and a $20\%$ test samples, in line with the standard cross validation procedure of machine learning models.
Our aim to compare the discrepancy of the predictions of either model against the true values in the test set. 
We will get three discrepancy measures from the ground truth: one for the LIN model, one for the LINS model, and one for the NNET model. 
The lowest discrepancy will determine the winning model.
As discrepancy measures we consider our whitened discrepancy measures and compare them with the commonly used euclidean distance (root mean squared error).
We will assume that the three response variables TASS, SNFD and TOVR are multivariate Gaussian, in line with their nature of continuous measurements.
In Table \ref{Performances}, we present the comparison of the predictive accuracy obtained with the three models, learned on the training set, and utilised to predict the true observations in the test set. 
In the first and second column, we present the root mean square error of the predictions (RMSE), calculated on the original response variables and on the normalised original data (obtained dividing each of the three response variables by their maximum value).
In the third and fourth column we present the whitened Wasserstein discrepancy (WASS, which, since we are assuming that the response variable are distributed as Gaussians, is equal to the whitened Fourier), for both the original and the normalised response.
In the fifth and sixth column we present the upper bound of the whitened Gini (GINI) discrepancy, again for both the original and the normalised response. 
For each discrepancy, we underline in bold the minimum value, which indicates the best model.

\begin{table}[!ht]
		\centering
		\footnotesize
		\begin{tabular}{cllllll}
			\hline
			& RMSEO & RMSEW & WASSO & WASSW & GINIO & GINIW \\
			\hline
			LIN & 552167 & 0.05235 & 2.94069 & 2.94069 & 3.82723 & 3.82723 \\ 
			LINS & \bf{544382} & \bf{0.05168} & 1.80616 & 1.80616 & 3.04339 & 3.04339  \\ 
			NNET & 554560 & 0.05352 & \bf{1.28701} & \bf{1.28701} & \bf{2.76701} & \bf{2.76701}\\  
			\hline
		\end{tabular}
		\caption{Comparison of predictive accuracies of the three considered models (LIN, LINS, NNET), with (W) and without (O) normalisation, using standard root mean squared error (RMSE), Whitened Wasserstein (WASS) and Whitened Gini upper bound (GINI).}
		\label{Performances}
	\end{table}

Table \ref{Performances} shows that the classical RMSE discrepancy varies under variable rescaling. 
In both cases, it leads to a minimum discrepancy for LINS, the multivariate regression model that explains financial variables with sustainability variables and the sector of belonging of the companies.
However, the values obtained by the RMSE vary, not only in absolute values, but also in relative values: the percentage advantage of LINS is about $12\%$ using the original variables and about $14\%$ using the normalised variables.
Differently, all our discrepancies lead to the same discrepancy values, regardless of whether the variables are normalised or not: a clear interpretational advantage.
In both cases, the LINS model has a lower discrepancy. 
For the Whitened Wasserstein discrepancy, the advantage of LINS with respect to LIN is about $28\%$.
For the Whitened Gini, the same advantage (in terms of the upper bound) is about $21\%$.
The previous results are consistent with what can be obtained applying a classical multivariate analysis of variance statistical test (MANOVA), which is applicable for multivariate Gaussian distributions as long as the models being compared are linear.  
In this case the application of the standard Pillai statistic (cf. Ref. \cite{Pillai}) gives a $p$-value equal to $0.003072$, which indicates rejecting the null hypotheses of a model without sectors against a model with sectors. 
We now consider the comparison between the multivariate linear models and the neural network. 
Such a comparison is not possible by means of standard statistical tests, such as MANOVA, as neural networks models are not linear and not nested with each other. 
For this reason, we resort a comparison conducted in terms of predictive accuracy. 
From this viewpoint, it is even more important to utilise a predictive accuracy measure that is invariant with respect to the measurement scale of the variables.
Table \ref{Performances} shows that, using both the whitened Wasserstein and the whitened Gini metrics, the neural network has the best performance. 
Furthermore, the value of the discrepancy is the same, for both the original and the normalised data. 
Differently, when we consider the commonly used Root mean squared error measure, based on the Euclidean distance, the neural network is worse than the linear model, and the value of the predictive accuracy depends on the measurement scale.
We thus infer that our proposed predictive accuracy metrics, based on the whitened discrepancies, do improve the standard metric, based on the mean squared error. 
Moreover, from an applied viewpoint, we conclude that Sustainability measures, in terms of ESG factors, affect company growth, in a non-linear manner, and depend on the activity sector of the companies.
For the sake of the interpretation, in Table \ref{Estimates} we report the estimated coefficients of the best linear model that we obtained using the original data in the training sample.
\begin{table}[!ht]
    \centering
    \footnotesize
    \begin{tabular}{llcc}
        \toprule
        Response & Explanatory variable & Estimate & p-value \\ 
        \hline
        TASS &\bf{Intercept} & -517579 & \bf{0.00017}\\ 
        TASS &			E.Sc &    213778  & 0.10923\\ 
        TASS &			\bf{S.Sc} &   189604 &  \bf{0.06006} \\ 
        TASS &			\bf{G.Sc} &   665923  & \bf{0.00002} \\ TASS & Financials &  -140600  & 
        0.45941   \\
        TASS &			Health \&Utilities & 84035  & 0.45311   \\ 
        TASS &   	Manufacturing & 49295    & 0.33215  \\
        TASS &			Tech. Com. & -112874  & 0.26517\\
        \hline
        SFND &\bf{Intercept}  & -185742  & \bf{0.00042} \\ 
        SFND &			E.Sc &   51961 & 0.30752   \\ 
        SFND &			\bf{S.Sc} &  109738 & \bf{0.00441}\\ 
        SFND &			\bf{G.Sc} &   240478   & \bf{0.00007}  \\ SFND & Financials &  -46898 & 
        0.51791 \\
        SFND &			Health \&Utilities & -1912  & 0.96431 \\ 
        SFND &   	Manufacturing & 16801    & 0.38646   \\
        SFND &			Tech. Com. & -47087 & 0.22331   \\
        \hline
        TOVR &\bf{Intercept}  & -289199  & \bf{0.01813}  \\ 
        TOVR &			E.Sc &    116485  & 0.32601 \\ 
        TOVR &			S.Sc &     92104 & 0.30384\\ 
        TOVR &			\bf{G.Sc} &  500069 & \bf{0.00039}  \\ TOVR & Financials &  -103705 & 
        0.53931 \\
        TOVR &		\bf{Health \&Utilities}	 & 358356 & \bf{ 0.00033} \\ 
        TOVR &   	Manufacturing & 11376    & 0.80119 \\
        TOVR &			Tech. Com. & -142515 & 0.11366\\
        \hline
    \end{tabular}
    \caption{Coefficient estimates of the linear model with sectors (LINS), learned on the training dataset. The significant coefficients are marked in bold.}
    \label{Estimates}
\end{table}

Table \ref{Estimates} shows that the Governance factor is positively correlated with company growth, in all its expressions: the higher the Governance score, the larger the company, in terms of total assets, equity and turnover.
The Social factor is also positively correlated with size, in terms of total assets and equity (which refer to a long term growth), but not in terms of turnover (which refers to a short term growth). 
The Environmental factor is, instead, not correlated with the size of companies, in line with the intuition that environmental scores depend more on the specific activity of a company, rather than on its size.
Finally, Table \ref{Estimates} indicates why the company sector affects the model performance: when a company belongs to the Health \& Utilities sector, it has an average increase in turnover of  $358,356$ euro, with respect to companies in the Consumer sector (which are estimated by the baseline intercept).  
We remark that, although neural networks are more accurate, as we have seen, they are also not explainable ``by design'', in terms of linear coefficients, as they are based on several estimated non linear coefficients whose interpretation is complicated, especially when the number of hidden nodes is increased. 
For explanation purposes, a multivariate linear model can be a good approximation to neural networks, recalling that a multivariate linear model is a neural network without hidden layers.
%


\section{Conclusion and perspectives}

In this paper, we introduced three new discrepancy measures to compare pairs of multivariate distributions. 
We show that all our proposed metrics are scale invariant and enjoy a generalized version of the uniform redistribution property. These two properties have a precise meaning in applied fields and thus are fundamental to comparing distributions meaningfully.
For example, the scale invariant property ensures that the discrepancy between two distributions describing two sets of data is not affected by the measuring unit used while gathering the data.
To complement our theoretical study of these discrepancies, we deploy them to study how sustainability factors affect the growth of companies and compare their performances with the results obtained from classic discrepancies, such as the Mean Square Error.
Our experiments showcase that scale invariant discrepancies have a clear interpretational advantage over classic methods. While the information gathered by the Mean Square Error depends on how we scale the data we study, the same does not hold for our discrepancies.
We believe that the range of applicability of such discrepancies is not limited to the study covered in this paper.
Among others,  this new way of approaching similarity between multivariate distributions can help to validate mathematical models through a better understanding of
big data toward predictive purposes on an ongoing observed dynamics.\cite{ABG}  
Further, these novel objects can be fruitfully used to evaluate the predictive accuracy of a machine learning model.
In this way, our discrepancies can be employed as a monitoring tool by the provider of a service/product based on artificial intelligence. 
These issues are defining problems in several different applied fields, ranging from health care, where comparing diagnostic tools in terms of their accuracy helps predict diseases, to text generation, where it is crucial to compare artificially generated texts in terms of their veridicality.  
Indeed, as the recently approved European Artificial Intelligence Act and all the discussion about AI regulation demonstrate, being able to effectively control and monitor the risks induced by AI-related products is a generation-defining challenge. We refer the reader also to a recent work in \emph{safe AI}, providing a Python toolbox.\cite{GudRaf2024}

In addition, since they treat effectively the problem of comparing multidimensional distributions in a unit-free fashion, the discrepancies we introduce could be useful tools for the \textcolor{blue}{\href{https://composite-indicators.jrc.ec.europa.eu/multidimensional-inequality}{Multidimensional Inequality Monitoring}} of the European Commission. This way, the inequality in the population with respect to, e.g., the joint distribution of wealth and education, can be compared among different European countries.

To summarize, we believe that our discrepancies can be used by service providers to improve the quality of their products as well as by authorities and policymakers.
\section*{Acknowledgment}

This work has been written within the activities of GNCS and GNFM groups of INdAM (Italian National Institute of High Mathematics). G.B.~has been funded by the European Union’s Horizon 2020 research and innovation programme under the Marie Sklodowska-Curie grant agreement No 101034413. P.G.  has been funded by the European Union - NextGenerationEU, in the framework of the GRINS- Growing Resilient, INclusive and Sustainable (GRINS PE00000018).




\begin{thebibliography}{99}
\bibitem{Aer} 
Aerts, S., Haesbroeck, G., and Ruwert, C., 
Multivariate coefficients of variation: comparison and influence functions, 
\textit{Journal of Multivariate Analysis}, \textbf{142} (2015)  183--198.

\bibitem{ABG}
Ajmone Marsan, G., Bellomo, N., and Gibelli, L., 
Stochastic evolutionary differential games toward a systems theory of behavioral social dynamics,
\emph{Math. Models Methods Appl. Sci.} \textbf{26} (6) (2016) 1051--1093. 

\bibitem{ambrosio2005gradient}
Ambrosio, L., Gigli, N, and Savar{\'e}, G. 
\emph{Gradient flows: in metric spaces and in the space of probability measures},
 (Springer Science \& Business Media, 2005).
 
\bibitem{Ana}
Ana Lugo, M., Comparing multidimensional indices of inequality: methods and application, Bishop, J. and Amiel, Y. (Ed.) Inequality and Poverty, Research on Economic Inequality, Vol. 14, (Emerald Group Publishing Limited, Leeds, 2007) 213--236.

\bibitem{Arnold}
Arnold, B.C., \textit{Pareto distributions}, (International Co-Operative Publishing House, Fairland, MD, USA 1983).

\bibitem{Au1}
Auricchio, G., Codegoni, A., Gualandi, S., Toscani, G., and Veneroni, M.,
On the equivalence between Fourier-based and Wasserstein metrics, \textit{Rendiconti Lincei. Matematica e Applicazioni}, \textbf{31}  (2020) 627--649.

\bibitem{Au2}
Auricchio, G., Codegoni, A., Gualandi, S., and Zambon, L.,
The Fourier discrepancy function, \textit{Communications in Mathematical Sciences}, \textbf{21}  (2023) 627--639.



\bibitem{AGT} 
Auricchio, G.,  Giudici, P., and Toscani, G., Extending the Gini index to higher dimension via whitening processes.
Preprint (2024)

\bibitem{GudRaf2024}
Babaei, G., Giudici, P. and Raffinetti, E., A Rank Graduation Box for SAFE artificial intelligence, \textit{Expert Systems with applications}, \textbf{259}  (2025)

\bibitem{barhen1995generalization}
Barhen, A. and Daudin, J.J.,
Generalization of the {M}ahalanobis distance in the mixed case,
  \textit{Journal of Multivariate Analysis}  \textbf{53} (2)  (1995)
  332--342
 
\bibitem{Ban}
Banerjee, S., Chakrabarti, B.K., Mitra, M. and Mutuswami, S.,
Inequality measures: the Kolkata index in comparison with other measures, \textit{Frontiers of Physics}, \textbf{8}  (2020) 562182.

\bibitem{BS}
Bell, A.J. and Sejnowski, T.J. , The independent components of natural scenes are edge filters, \emph{Vision Research} \textbf{37}  (1997) 3327--3338.

\bibitem{BL}
Betti, G., and Lemmi, A., \textit{Advances on income inequality and concentration measures}, (Routledge, New-York  2008).

\bibitem{BCT}
Bisi,  M., Carrillo, J.A. and Toscani, G., Decay rates in probability metrics towards homogeneous cooling states for the inelastic Maxwell model,  \emph{Journal of Statistical Physics}, \textbf{124} (2-4) (2006) 625--653.

\bibitem{brenier1991polar}
Brenier, Y.,
Polar factorization and monotone rearrangement of vector-valued functions,
  \textit{Communications on pure and applied mathematics},
 \textbf{44}, (4)  (1991) 375--417.
 
\bibitem{CCG}
 Carlen, E.A.,   Carvalho, M.C., and Gabetta, E., Central limit theorem for Maxwellian molecules and truncation of the Wild expansion. \emph{Communications on Pure and Applied Mathematics}, \textbf{53} (3) (2000) 370--397.
 
 \bibitem{CGT}
Carlen, E.A.,    Gabetta, E., and Toscani, G., Propagation of smoothness and the rate of exponential convergence to equilibrium for a spatially homogeneous Maxwellian gas,
\emph{Communications in mathematical physics}, \textbf{199} (3) (1999) 521--546.
 

\bibitem{CaTo}
Carrillo, J.A. and Toscani, G., Contractive probability metrics and asymptotic behavior of dissipative kinetic equations, \textit{Rivista di Matematica dell'Universit\`a di Parma}, \textbf{6} (2007) 75--198.

\bibitem{chafai2010}
 Chafa\"i, D.,  
    {Wasserstein distance between two {G}aussians},\\
  {https://djalil.chafai.net/blog/2010/04/30/wasserstein-distance-between-two-gaussians/}, (2010).
  

\bibitem{Cou}
Coulter, P.B., \textit{Measuring inequality: A methodological handbook}, (Westview Press, Boulder,  1989).

\bibitem{Cow} 
Cowell, F., \textit{Measuring inequality}, (Oxford University Press, Oxford,   2011).

\bibitem{cuturi2013sinkhorn}
Cuturi, M., Sinkhorn distances: {L}ightspeed computation of optimal transport, \textit{Advances in neural information processing systems} \textbf{26}  (2013).

\bibitem{Decancq}
Decancq, K. and Ana Lugo, M., Inequality of wellbeing: A multidimensional approach, \textit{Economica}, \textbf{79}  (2012) 721--746.

\bibitem{DelGiudice1}
Del Giudice, M., Heterogeneity coefficients for Mahalanobis' $D$ as a multivariate effect size, \textit{Multivariate Behavioural Research}, \textbf{52}  (2017) 216--221.

\bibitem{DelGiudice2}
Del Giudice, M., Addendum to: heterogeneity coefficients for Mahalanobis' $D$ as a multivariate effect size,  \textit{Multivariate Behavioural Research}, \textbf{53}  (2018) 571--573.

\bibitem{Eli}
Eliazar, I.,  A tour of inequality, \textit{Annals of Physics}, \textbf{389}  (2018) 306--332.

\bibitem{Eli3}
Eliazar, I. and Giorgi, G.M.,  From Gini to Bonferroni to Tsallis: an inequality-indices trek, \textit{Metron}, \textbf{78}  (2020) 119--153.

\bibitem{evans1999differential}
Evans, L.C. and Gangbo, F.,
  \emph{Differential equations methods for the {M}onge-{K}antorovich mass transfer problem},
  (American Mathematical Soc.,   1999)


\bibitem{Fri} 
Friedman, J.H.,  Exploratory Projection Pursuit, \emph{Journal of the American Statistical Association}, \textbf{82} (1987) 249--266.

\bibitem{GTW}
Gabetta, E., Toscani, G., and Wennberg, B., Metrics for probability measures and the trend to equilibrium for solutions of the Boltzmann equation, \textit{Journal of Statistical Physics}, \textbf{81}  (1995) 901--934.

\bibitem{Gini1}
Gini, C.,  Sulla misura della concentrazione e della variabilit\`a dei caratteri, \textit{Atti del Reale Istituto Veneto di Scienze, Lettere ed Arti}, \textbf{73}  (1914) 1203--1248. English translation in \textit{Metron}, 3--38, (2005).

\bibitem{Gini2}
Gini, C.,  Measurement of inequality of incomes, \textit{The Economic Journal}, \textbf{31}  (1921) 124--126.


\bibitem{GudRaf2023}
Giudici, P. and Raffinetti, E., SAFE Artificial Intelligence in Finance, \textit{Finance Research Letters}, \textbf{56}  (2023).


\bibitem{giudici2024measuring}
Giudici, P., Raffinetti, E,, and Toscani, G.,
  {Measuring multidimensional inequality: a new proposal based on the {F}ourier transform}'',
 {arXiv preprint arXiv:2401.14012}  (2024).
 
 \bibitem{GJT}
 Goudon, T.,  Junca, S. and Toscani, G., Fourier-based distances and Berry-Esseen like inequalities for smooth densities, \emph{Monatshefte f\"ur Mathematik}, \textbf{135} (2)  (2002) 115--136.
 
\bibitem{Grothe}
Grothe, O., K{\"a}Kele, F. and Schmid, F.,  A multivariate extension of the Lorenz curve based on copulas and a related multivariate Gini coefficient, \textit{The Journal of Economic Inequality}, \textbf{20} (2022) 727--748.

\bibitem{HN} 
Hao, L. and Naiman, D.Q., \textit{Assessing inequality}, (Sage, Los Angeles, 2010).

\bibitem{HR}
Hurley, N., and Rickard, S., Comparing measures of sparsity, \textit{IEEE Transactions on Information Theory}, \textbf{55},  (2009) 4723--4741. 

\bibitem{kantorovich1960mathematical}
Kantorovich, L.V., 
Mathematical methods of organizing and planning production, 
 \textit{Management science}, \textbf{6} (4)  (1960)
  366--422.

\bibitem{KLS}
Kessy, A.,  Lewin, A. and  Strimmer, K.,   Optimal whitening and decorrelation, 
\emph{The American Statistician} \textbf{72} (4) (2018) 309--314.


\bibitem{KoshMos96}
Koshevoy, G. and Mosler, K.,  The Lorenz zonoid of a multivariate distribution, \textit{Journal of the American Statistical Association}, \textbf{91}  (1996) 873--882

\bibitem{KoshMos97} 
Koshevoy, G. and Mosler, K.,  Multivariate Gini indices, \textit{Journal of Multivariate Analysis}, \textbf{60}  (1997) 252--276.

\bibitem{LZ} 
Li, G. and Zhang, J., Sphering and its properties, \emph{Sankhya A} \textbf{60} (1998)  119--133. 

\bibitem{Lor}
Lorenz, M.,  Methods of measuring the concentration of wealth, \textit{Publications of the American Statistical Association}, \textbf{9}  (1905)  209--219. 

\bibitem{Maha}
Mahalanobis, P.C.,  On the generalised distance in statistics, in \textit{Proceedings of the National Institute of Sciences of India}, \textbf{2}  (1936) 49--55 (Retrieved 2016-09-27).

\bibitem{monge1781memoire}
Monge, G.,
  {M{\'e}moire sur la th{\'e}orie des d{\'e}blais et des remblais},
  \textit{Mem. Math. Phys. Acad. Royale Sci.}, (1781)
  {666--704}.

\bibitem{peyre2019computational}
Peyr\'e, G. and Cuturi, M., Computational optimal transport: {W}ith applications to data science, \textit{Foundations and Trends in Machine Learning} \textbf{11} (5-6) (2019) 355--607. 

\bibitem{Pie}
Pietra, G.,  Delle relazioni tra gli indici di variabilit\`a. Nota I, \textit{Atti del Reale Istituto Veneto di Scienze, Lettere ed Arti}, \textbf{74}  (1915) 775--804.

\bibitem{Pillai}
Pillai, K.C.S., Some New test criteria in multivariate analysis, \emph{Annals of mathematical statistics}, \textbf{26} (1), (1955) 117--121.
     
\bibitem{PT}
Pulvirenti, A. and Toscani, G., Asymptotic properties of the inelastic Kac model, \emph{Journal of Statistical Physics}, \textbf{114} (5-6) (2004) 1453--1480.

\bibitem{otto2000generalization}
Otto, F. and Villani, C.,
Generalization of an inequality by {T}alagrand and links with the logarithmic {S}obolev inequality, 
\textit{Journal of Functional Analysis}, \textbf{173} (2) (2000) 361--400.

\bibitem{santambrogio2015optimal}
 Santambrogio, F.,   
 \textit{Optimal transport for applied mathematicians},
  Vol. 55, ( Birk{\"a}user, NY, 2015).
  
  \bibitem{Sarabia}
Sarabia, J.M. and Jorda, V.,  Lorenz surfaces based on the Sarmanov–Lee distribution with applications to multidimensional inequality in well-being, \textit{Mathematics}, \textbf{8}  (2020) 2095.


\bibitem{Tagushi1}
Taguchi, T.,  On the two-dimensional concentration surface and extensions of concentration coefficient and Pareto distribution to the two-dimensional case-I, \textit{Annals of the Institute of Statistical Mathematics}, \textbf{24} (1972) 355--382.

\bibitem{Tagushi2}
Taguchi, T.,  On the two-dimensional concentration surface and extensions of concentration coefficient and Pareto distribution to the two-dimensional case-II, \textit{Annals of the Institute of Statistical Mathematics}, \textbf{24}  (1972) 599--619.

\bibitem{ToTo}
Torregrossa, M. and Toscani, G.,  Wealth distribution in presence of debts. A Fokker-Planck description, \textit{Communications in Mathematical Sciences}, \textbf{16} (2018) 537--560.

\bibitem{To1}
Toscani, G.,  On Fourier-based inequality measures, \textit{Entropy}, \textbf{24}  (2022) 1393.

\bibitem{To2}
Toscani, G., Measuring multidimensional heterogeneity in emergent social phenomena, \textit{European Journal of Applied Mathematics} (in press)  (2024).

\bibitem{TV}
Toscani, G. and Villani, C., Probability metrics and uniqueness of the solution to the Boltzmann equation for a Maxwell gas, \emph{Journal of Statistical Physics}, \textbf{94} (3-4) (1999) 619--637.

\bibitem{villani2009optimal}
Villani, C.,
 \textit{Optimal transport: old and new},
  Vol. {338}, (Springer-Verlag, Basel, 2009).

\bibitem{VN}
Voinov, V.G. and Nikulin M.S., \textit{Unbiased estimators and their applications, 2, multivariate case}, (Kluwer, Dordrecht,  1996).

\bibitem{Zol} 
Zolotarev, V.M.,  One-dimensional stable distributions, \textit{Translations of Mathematical Monographs}, \textbf{65}  ( American Mathematical Society, Providence, 1986).


\bibitem{ex1}
Hao, Ning, Bin Dong, and Jianqing Fan. Sparsifying the Fisher linear discriminant by rotation. \emph{Journal of the Royal Statistical Society Series B: Statistical Methodology} \textbf{77} (4) (2015) 827--851.

\bibitem{ex2}
Zuber, V., and Korbinian S., Gene ranking and biomarker discovery under correlation. \emph{Bioinformatics} \textbf{25} (20) (2009) 2700--2707.

\end{thebibliography}
\end{document}